\documentclass[number,12pt,sort&compress,final]{elsarticle}
\makeatletter
\def\ps@pprintTitle{%
 \let\@oddhead\@empty
 \let\@evenhead\@empty
 \def\@oddfoot{}%
 \let\@evenfoot\@oddfoot}
\makeatother

\usepackage{graphicx}
\usepackage{xcolor}
\usepackage{amssymb}
\usepackage{mathtools}
\usepackage{mathrsfs}
\usepackage{braket}
\usepackage{graphicx}
\usepackage{subfig}
\usepackage{soul}
\usepackage{xcolor}
\usepackage[english]{babel}
\usepackage{blindtext}
\usepackage{eqnarray}
\usepackage{dsfont}
\usepackage{ulem}

\DeclareFontFamily{OT1}{pzc}{}
\DeclareFontShape{OT1}{pzc}{m}{it}{<-> s * [1.2] pzcmi7t}{}
\DeclareMathAlphabet{\mathpzc}{OT1}{pzc}{m}{it}
\newcommand{\rme}[0]{\mathrm{e}}
\newcommand{\rmi}[0]{\mathrm{i}}
\newcommand{\rmd}[0]{\mathrm{d}}
\newcommand{\Xvec}[0]{\mathbf{X}}
\newcommand{\XL}[0]{X_{\textsc{l}}}
\newcommand{\PL}[0]{P_{\textsc{l}}}
\newcommand{\XM}[0]{X_{\textsc{m}}}
\newcommand{\PM}[0]{P_{\textsc{m}}}
\newcommand{\Xv}[0]{X_{\textsc{v}}}
\newcommand{\Pv}[0]{P_{\textsc{v}}}
\newcommand{\XLo}[0]{X_{\textsc{l}\scriptscriptstyle{1}}}
\newcommand{\PLo}[0]{P_{\textsc{l}\scriptscriptstyle{1}}}
\newcommand{\XLt}[0]{X_{\textsc{l}\scriptscriptstyle{2}}}
\newcommand{\PLt}[0]{P_{\textsc{l}\scriptscriptstyle{2}}}
\newcommand{\XMo}[0]{X_{\textsc{m}\scriptscriptstyle{1}}}

\newcommand{\XMt}[0]{X_{\textsc{m}\scriptscriptstyle{2}}}

\newcommand{\PVo}[0]{P_{\textsc{v}\scriptscriptstyle{1}}}

\newcommand{\PVt}[0]{P_{\textsc{v}\scriptscriptstyle{2}}}

\begin{document}
\begin{frontmatter}

\title{Generating mechanical and optomechanical entanglement via pulsed interaction and measurement \vspace{-0.25cm}}
\author[1,4,5]{J.~Clarke}
\author[1,4]{P.~Sahium}
\author[1]{K.~E.~Khosla}
\author[2,3]{I.~Pikovski}
\author[1]{M.~S.~Kim}
\author[1,5]{M.~R.~Vanner}
\address[1]{QOLS, Blackett Laboratory, Imperial College London, London SW7~2AZ, UK}
\address[2]{The Oskar Klein Centre, Department of Physics, Stockholm University, Stockholm 106~91, Sweden}
\address[3]{Department of Physics, Stevens Institute of Technology, Hoboken, NJ~07030, USA \vspace{-0.75cm}}
\fntext[4]{These authors contributed equally.}\fntext[5]{Email: jack.clarke@imperial.ac.uk, m.vanner@imperial.ac.uk}
\begin{abstract}
{\small{Entanglement generation at a macroscopic scale offers an exciting avenue to develop new quantum technologies and study fundamental physics on a tabletop. Cavity quantum optomechanics provides an ideal platform to generate and exploit such phenomena owing to the precision of quantum optics combined with recent experimental advances in optomechanical devices.
In this work, we propose schemes operating outside the resolved-sideband regime, to prepare and verify both optical-mechanical and mechanical-mechanical entanglement. 
Our schemes employ pulsed interactions with a duration much less than the mechanical period and, together with homodyne measurements, can both generate and characterize these types of entanglement.
To improve the performance of our schemes, a precooling stage comprising prior pulses can be utilized to increase the amount of entanglement prepared, and local optical squeezers may be used to provide resilience against open-system dynamics.
The entanglement generated by our schemes is quantified using the logarithmic negativity and is analysed with respect to the strength of the pulsed optomechanical interactions for realistic experimental scenarios including mechanical decoherence and optical loss. 
Two separate schemes for mechanical entanglement generation are introduced and compared: one scheme based on an optical interferometric design, and the other comprising sequential optomechanical interactions.
The pulsed nature of our protocols provides more direct access to these quantum correlations in the time domain, with applications including quantum metrology and tests of quantum decoherence. By considering a parameter set based on recent experiments, the feasibility to generate significant entanglement with our schemes, even with large optical losses, is demonstrated. 
}}
\end{abstract}

\end{frontmatter}
\section{Introduction}
\label{sec:Intro}

Entanglement is one of most striking features of quantum mechanics and allows for correlations between two or more objects to be much stronger than is allowed classically. Though this behaviour troubled the founders of quantum mechanics~\cite{EPR1935,schrodinger1935gegenwartige}, entanglement is now very well established, and is viewed as a powerful resource for quantum technologies, such as quantum metrology~\cite{giovannetti2011advances} and communication~\cite{gisin2002quantum}, and for tests of fundamental physics~\cite{bell1964einstein}.

The non-classical correlations of entanglement have now been observed in many of the facets of quantum optics including light with continuous~\cite{heidmann1987observation, furusawa1998unconditional, Braunstein2005} and discrete~\cite{Freedman1972, aspect1981, walther2004broglie, mitchell2004super} degrees-of-freedom, as well as the electronic states of neutral atoms~\cite{Julsgaard2001, Chou2007,hofmann2012heralded} and nitrogen-vacancy centres~\cite{Bernien2013}.

Entanglement is now also being actively studied both theoretically and experimentally for motional degrees of freedom, opening a rich new avenue for quantum science. Notably, the first demonstration of entanglement between spatially distinct mechanical oscillators has been achieved using trapped ions~\cite{Jost2009}, which reported the observation of non-classical correlations between the motion of two separated pairs of ions. One route to study motional entanglement for more massive systems is via cavity quantum optomechanics~\cite{meystre2013short,aspelmeyer2014cavity}, which utilizes radiation-pressure and other optical forces, such as electrostriction, to generate and study non-classical motional states of mechanical oscillators from the zeptogram to kilogram scale.

Studying the various forms of optomechanical entanglement is an active current area of research and there have been several theoretical, and more recently experimental, studies exploring this avenue. While we do not aim to provide a thorough review here, we briefly describe and contrast some key studies of both optical-mechanical and  mechanical-mechanical entanglement. 

Focusing first on theoretical proposals, a notable early direction on optomechanical entanglement was to use photon-phonon transfer operations~\cite{Parkins1999}, where non-classical and entangled states of light are mapped onto the motion of mechanical oscillators via light-matter beamsplitter interactions~\cite{Zhang2003}. Interactions of this type require operation in the resolved-sideband regime of optomechanics, which is realized if the mechanical frequency far exceeds the cavity decay rate. Following these proposals, detailed studies into the generation and detection of Gaussian optical-mechanical entanglement have been carried out~\cite{vitali2007optomechanical,paternostro2007creating}, which focus on the linearized dynamics around the steady state. In the steady state, optomechanical systems are subject to certain stability conditions, which preclude parts of parameter space, particularly for optical drive on the blue sideband. Long-pulsed optical drives operating in the resolved-sideband regime were suggested as a means to access this part of parameter space~\cite{Hofer2011}, and the effect of high thermal occupation~\cite{Rakhubovsky2015}, multiple interactions, pulse shaping, and different optical detunings have also been studied for such long optical pulses~\cite{Lin2015}.

In addition to optical-mechanical entanglement, developing methods to establish entanglement between a pair of mechanical oscillators is also of key interest. In particular, the steady-state linearized dynamics of two mechanical oscillators interacting with the same entangling optical field have been studied to investigate Gaussian entanglement~\cite{mancini2002entangling}. Alternative experimental configurations have also been proposed, such as the suspension of two mechanical membranes within the same optical cavity~\cite{hartmann2008steady}, and the injection of squeezed light into both a double cavity system~\cite{pinard2005entangling} and a ring cavity design~\cite{huang2009entangling}. The generation of mechanical entanglement via reservoir engineering of a single cavity mode by a multi-tone drive has also been suggested~\cite{woolley2014}. Recently, a more comprehensive analysis of steady-state Gaussian entanglement has been carried out, which investigates optical-mechanical, mechanical-mechanical, and tripartite light-mechanics entanglement~\cite{brunelli2019}. Importantly, this analysis goes beyond the usual resolved-sideband regime and does not employ the rotating-wave or adiabatic approximations typically used in the literature.

Swapping optical-mechanical to mechanical entanglement by long-pulsed interactions, optical interferometry, and measurements provides another exciting route for generating entangled states of mechanical oscillators. In this setting, linear interactions and measurements on the optical field have been discussed as a way to generate Gaussian continuous-variable entanglement~\cite{pirandola2006macroscopic,Vostrosablin2016}. Furthermore, complementary approaches that implement photon-counting measurements to generate and witness non-Gaussian mechanical entanglement have also been considered~\cite{vacanti2008,borkje2011}. Additionally, recent theoretical work, which proposes entanglement swapping between two optical and two mechanical modes offers a promising avenue to use mechanical entanglement to investigate spatially dependent decoherence in massive systems~\cite{kiesewetter2017pulsed}. 

In recent years, field-mechanics and mechanics-mechanics entanglement experiments have been performed, demonstrating the interest in, and feasibility of, generating optomechanical entanglement. Notably, long-pulsed optomechanical interactions in the resolved-sideband regime have allowed entanglement between microwave fields and mechanical motion to be measured using optical-quadrature measurements~\cite{palomaki2013}. While in the optical domain, photon counting measurements have been used to witness non-classical optomechanical correlations between optical pulses and phonon modes in bulk diamond via off-resonant Raman scattering~\cite{lee2012macroscopic}, and then similarly in a nanomechanical resonator operating in the resolved-sideband regime~\cite{riedinger2016}. Furthermore, mechanical entanglement has been established using an interferometric pump-probe scheme, first between two spatially-separated diamonds with photon-counting measurements on Stokes scattered photons~\cite{lee2011entangling}, and then between mechanical resonators in the resolved-sideband regime of cavity optomechanics~\cite{riedinger2018}. Moreover, continuous driving of a non-interferometric configuration with two micromechanical drum oscillators has allowed for mechanical entanglement to be established mediated by microwave fields~\cite{ockeloen2018}.
    
In this work, we propose methods for preparing and verifying both optical-mechanical and mechanical-mechanical entanglement that explore new parameter regimes in the landscape of optomechanical entanglement. Our protocols utilize short optical pulses with a temporal width much less than the mechanical period, such that over the optomechanical interaction both mechanical evolution and decoherence can be ignored. Pulsed optomechanics~\cite{Vanner2011} has allowed for the development of numerous non-classical state preparation and verification schemes, see e.g. Refs~\cite{Buchmann2012,Sekatski2014,hoff2016,ringbauer2018,clarke2019}, by utilizing resonant interactions in the unresolved-sideband regime---where the cavity decay rate greatly exceeds the mechanical frequency. This regime of optomechanics also allows the optical and mechanical quadratures to become strongly correlated over a pulsed interaction, which provides a route for generating optomechanical entanglement. We consider pulsed-linearized interactions that enable the generation and full characterization of Gaussian continuous-variable entangled states, which allows for greater entanglement than is possible in low-dimensional discrete variable states. The pulsed nature of our protocols also allows for more direct access to the entanglement in the time domain and permits operation in discrete time steps. Additionally, we compare and contrast interferometric and non-interferometric configurations and analyze the utility of optical squeezing in each case. Here, we see that optical squeezing provides a route to increase the resilience of the entangled state to optical loss and mechanical decoherence. 

Both long-pulsed and short-pulsed optomechanical interactions present promising and distinct routes for exploring entanglement generation in the laboratory. In particular, note that the effective Hamiltonians describing the short-pulsed and long-pulsed interactions are different. In the latter case, the bare optomechanical-interaction is integrated over an interval where there is appreciable mechanical evolution. Typically, this difference in the duration of the interaction means that long pulses are used to achieve beamsplitter or two-mode squeezer interactions when the drive is detuned, whereas short pulses couple the optical amplitude and mechanical position during an interval in which there is negligible free mechanical evolution. Furthermore, a consequence of the different requirements on the system parameters, in particular the cavity decay rate and the mechanical resonance, is that the two regimes are relevant to very different physical setups.

Experimental progress in optomechanics outside the resolved-sideband regime has advanced to a point where the entanglement schemes we propose are feasible with current systems. For example, mechanical cooling via pulsed-interaction and measurement has been performed~\cite{vanner2013cooling,muhonen2019state}, and developments in sliced-photonic crystal structures now enable large optomechanical coupling rates~\cite{leijssen2015strong,leijssen2017nonlinear}. Moreover, it has been experimentally demonstrated that pulsed interactions can provide resilience to optics-induced thermal heating effects owing to the slow thermal timescale~\cite{Meenehan2015}. These experimental advances allow our pulsed optical-mechanical entanglement protocol to be realized with present-day parameters, and our approach to mechanical-mechanical entanglement can be achieved even with total optical efficiencies much less than unity, i.e. of order $10\%$. 

Further work in this direction has a wide range of applications for quantum technologies and fundamental science. To name a few, investigating optomechanical entanglement in the unresolved-sideband regime using short optical pulses provides new routes to: develop quantum sensors~\cite{meystre2013short} and quantum parameter estimation techniques~\cite{Zheng2016}, perform optical-to-mechanical state swaps~\cite{Bennett2016} and optomechanical transduction~\cite{vostrosablin2018}, measure decoherence rates in massive systems~\cite{kiesewetter2017pulsed}, enable entanglement distillation via optomechanics~\cite{Montenegro2019}, investigate non-classical correlations mediated by gravity~\cite{bose2017spin,marletto2017gravitationally,miao2019}, and place new bounds on free parameters in generalized uncertainty principle models by exploring optomechanical geometric phases~\cite{Pikovski2012,Khosla2013,Armata2016,Bosso2017}.

In Section \ref{pulsed om}, we describe the physical operations that are used in our protocols for optomechanical entangled state preparation and verification. In particular, we describe linearized pulsed-optomechanical interactions, thermal decoherence channels, and optical homodyne measurements. We do so by using the covariance matrix description of quantum states in phase space to efficiently calculate the effect of these Gaussian processes. In Section \ref{light-mechanics}, we use these operations as building blocks to develop a protocol for the preparation and verification of entanglement between light and mechanics. In Section \ref{twomechanics}, we then build upon these ideas to introduce two protocols that exploit this optical-mechanical entanglement to generate entanglement between two mechanical oscillators.

\section{Pulsed interactions and measurements} \label{pulsed om}

\subsection{Pulsed optomechanics and Gaussian operations}
For a pulsed optomechanical interaction~\cite{Vanner2011}, all the optical timescales involved are much less than the mechanical period, and the pulse is accommodated by operating in the unresolved-sideband regime, where the cavity decay rate $\kappa$ far exceeds the mechanical angular frequency $\omega_{\textsc{m}}$. We consider an optomechanical system comprising a cavity light mode and a mechanical mode interacting via radiation pressure. In a frame rotating at the cavity frequency, the optomechanical interaction is described by the Hamiltonian 
\begin{align} 
    H = \hbar \omega_{\textsc{m}} b^{\dagger} b - \hbar g_0 a^{\dagger} a (b+b^{\dagger}),
\label{hamiltonian}
\end{align}
where $g_{0}$ is the intrinsic optomechanical coupling rate~\cite{law1995}. Here, $a$ and $b$ are the optical and mechanical annihilation operators, respectively, and we define the dimensionless optical quadratures as $X_{\mathrm{L}} = (a + a^{\dagger})/\sqrt{2}$ and $P_{\mathrm{L}} = - \rmi(a - a^{\dagger})/\sqrt{2}$, which satisfy the canonical commutation relation $[{X_{\mathrm{L}}},{P_{\mathrm{L}}}] = \mathrm{i}$. The mechanical quadratures $X_{\mathrm{M}}$ and $P_{\mathrm{M}}$ are defined in the same way using the mechanical annihilation and creation operators. 

The optomechanical Hamiltonian is linearized by transforming to a displaced frame such that $a\rightarrow\alpha+a$---where $\alpha$ is the mean intracavity amplitude of the pulse---and neglecting terms quadratic in $a$ and $a^{\dagger}$ that describe the small quantum fluctuations. This linearization leads to
\begin{align} 
    H_{\textsc{lin}} = \hbar \omega_{\textsc{m}} b^{\dagger} b - \hbar g_{0}\alpha^2(b+b^{\dagger}) -\hbar g_{0}\alpha(a+a^{\dagger})(b+b^{\dagger}),
\label{hamiltonianlinear}
\end{align}
where we have taken $\alpha\in\mathds{R}$ without loss of generality. In the unresolved-sideband regime, we describe a pulsed-optomechanical interaction with the unitary operator $U_{\textsc{lin}} = \mathrm{e}^{\rmi \lambda X_{\mathrm{M}}} \mathrm{e}^{\rmi \chi X_{\mathrm{L}} X_{\mathrm{M}}}$. The first exponential term in $U_{\textsc{lin}}$ represents a deterministic momentum transfer to the mechanics due to the mean photon number, where $\lambda\propto\alpha^2{g}_{0}/\kappa$ quantifies this transfer in units of zero-point momentum. The second exponential in $U_{\textsc{lin}}$ is a momentum transfer to the mechanical mode dependent upon the amplitude quadrature of the light, where the optomechanical interaction strength is given by $\chi\propto\alpha{g}_{0}/\kappa$. The constants of proportionality in front of $\lambda$ and $\chi$ are of order unity and depend on the optical pulse shape. From here on, we will use $U_{\textsc{om}} = \rme^{\rmi \chi \XL\XM}$ to describe the optomechanical interaction, as the deterministic part $\rme^{\rmi \lambda \XM}$ may either be compensated for by applying a suitable displacement or subtracted from the measurement results in postprocessing.

As the linearized optomechanical Hamiltonian is bilinear in annihilation and creation operators, the unitary it generates will transform Gaussian states into other Gaussian states~\cite{Weedbrook2011} . Such states can be fully characterized by the first and second moments of their quadrature vector $\Xvec=(\mathbf{\XL},\mathbf{\XM})^{\mathrm{T}}=(\XL,\PL,\XM,\PM)^{\mathrm{T}}$. Here, the quadrature vectors for each mode are given by $\mathbf{\XL}=(\XL,\PL)^{\mathrm{T}}$ and  $\mathbf{\XM}=(\XM,\PM)^{\mathrm{T}}$, respectively, and the canonical commutation relations can be written as $[\Xvec,\Xvec^{\mathrm{T}}]=\rmi\Omega$, where the symplectic form $\Omega$ corresponding to our chosen ordering of quadrature operators, is in general, given by
\begin{equation}
    \Omega=\bigoplus_{i=1}^n\begin{pmatrix}0&1\\-1&0\end{pmatrix},
\end{equation}
where $n=2$ for the bipartite optomechanical system. The first moments are given by the expectation value of the quadrature vector, $\braket{\Xvec}$, while the second moments correspond to elements of the covariance matrix $\sigma$, which are given by
\begin{align}
    \sigma_{i,j} = \frac{1}{2} \langle X_{i}X_{j}+X_{j}X_{i} \rangle - \langle {X_i} \rangle \langle {X_j} \rangle.
\end{align}
In this work, we utilize the block matrix form of the covariance matrix:
\begin{align}
    \sigma = \begin{pmatrix}
						A & C \\ 
						C^\mathrm{T} & B
					 \end{pmatrix}.
\label{blockCM}
\end{align}
For a system comprising one optical and one mechanical mode, the $A$, $B$, and $C$ blocks are $2\times2$ matrices, where $A$ and $B$ correspond to the optical and mechanical modes, respectively, and the off-diagonal block $C$ corresponds to optomechanical correlations.

The set of Gaussian operations we employ in this work may be divided into unitary and non-unitary operations. A general Gaussian unitary on $n$ modes can be written with a symplectic matrix $S{\in}Sp(2n,\mathds{R})$ acting on the quadrature vector $\Xvec$, where the real symplectic group $Sp(2n,\mathds{R})$ is defined as the group of $2n\times2n$ real symmetric matrices that preserve the commutation relations between the quadrature operators, namely $S\Omega{S}^\mathrm{T}=\Omega$. If the action of some Gaussian unitary $U_{\textsc{G}}$ on a state $\rho$ in Hilbert space is given by $U_{\textsc{G}}\rho{U}_{\textsc{G}}^{\dagger}$, then in quantum phase space the mapping is described by $\Xvec\rightarrow{S}\Xvec$, where $S$ is the symplectic matrix that has the same action on the quadrature vector as does the Gaussian unitary in the Heisenberg picture. For later convenience, we use the circle notation for quantum operations on density matrices, i.e. $M\circ\rho=M\rho{M}^{\dagger}$.

On the other hand, non-unitary Gaussian operations---optical loss, mechanical decoherence and optical homodyne measurements---are described differently. We model the decoherence of each mode using a phase-insensitive Gaussian channel, which corresponds to each mode interacting with a Gaussian environment on a beamsplitter~\cite{Holevo2001, Paris2005}. In Hilbert space we represent the action of these phase-insensitive Gaussian decoherence channels on state $\rho$ as $\mathcal{E}_{\eta}(\rho)$ and $\mathcal{E}_{\gamma}(\rho)$, where $\eta$ and $\gamma$ are the efficiency of the optical channel and the damping rate of the mechanical mode, respectively. As we work in the pulsed regime of optomechanics, the mechanical decoherence may be neglected over the course of the optomechanical interactions and optical measurements.

Gaussian measurements may be described using the Krauss operator representation, i.e. $\rho\propto K_{\textsc{g}}\circ\rho$, where $K_{\textsc{g}}$ is the Krauss operator corresponding to the Gaussian measurement. Also, by considering the first-moments vector $\braket{\textbf{X}}$ and the covariance matrix of Eq.~\eqref{blockCM}, a Gaussian measurement on the optical mode can be described through the mapping~\cite{Paris2005}
\begin{align}
\label{MeasEquation}
    \braket{\mathbf{\XM}} &\rightarrow\braket{\mathbf{\XM}}+C^{\mathrm{T}}(A+\sigma_{\mathrm{meas}})^{-1}(\braket{\Xvec_{\mathrm{meas}}}-\braket{\mathbf{\XL}}),\\
    B &\rightarrow{B}-C^{\mathrm{T}}(A+\sigma_{\mathrm{meas}})^{-1}C\label{covMeasEquation}.
\end{align}
where $\braket{\textbf{X}_{\mathrm{meas}}}$ corresponds to the measurement outcome and $\sigma_{\mathrm{meas}}$ describes the Gaussian measurement~\cite{eisert2002distilling}.

Throughout this paper we describe quantum states as either density matrices in Hilbert space or as covariance matrices in phase space. We use these descriptions interchangeably and the corresponding symplectic transformations, together with unitary and non-unitary operations, can be found in \ref{symplectictransformations}. 

\subsection{Precooling and optomechanical entanglement}
The symplectic formalism outlined above may then be used to efficiently compute the generation of optical-mechanical entanglement and also be used to describe the effects of optical-quadrature measurements. Moreover, following a pulsed optomechanical interaction with an optical measurement allows one to reduce thermal noise along an axis of mechanical phase space, which we refer to as mechanical precooling.

The generation of an entangled optomechanical state is achieved via interaction of light and mechanics through $U_{\textsc{om}}$. However, the optical mode will then be subject to optical loss described by the channel $\mathcal{E}_{\eta}$, and we assume that the optical environment is well described by the vacuum state. After this optical loss channel, the mechanical state evolves freely through a phase-space angle $\theta$, described by the unitary ${U}_{\textsc{rot}}(\theta)=\rme^{\rmi\theta{b}^{\dagger}b}$. During this time, the mechanical mode interacts with a thermal environment with mean occupation $\bar{N}$ at a rate $\gamma$, described by $\mathcal{E}_{\gamma}$. Including this decoherence, the map that describes the generation of an optical-mechanical entangled state from initial state $\rho$ is therefore
\begin{equation}\label{omint}
     \rho\rightarrow\mathcal{E}_{\gamma}({U}_{\textsc{rot}}(\theta)\circ\mathcal{E}_{\eta}({U}_{\textsc{om}}\circ\rho)).
\end{equation}
Note that the phase insensitive channel commutes with the free evolution of the mechanical mode, i.e. $\mathcal{E}_{\gamma}({U}_{\textsc{rot}}\circ\rho)={U}_{\textsc{rot}}\circ\mathcal{E}_{\gamma}(\rho)$.

\begin{figure}
\vspace{-2cm}\hspace{-1cm} %%was width=1.15\linewidth
\includegraphics[width=1.1\linewidth]{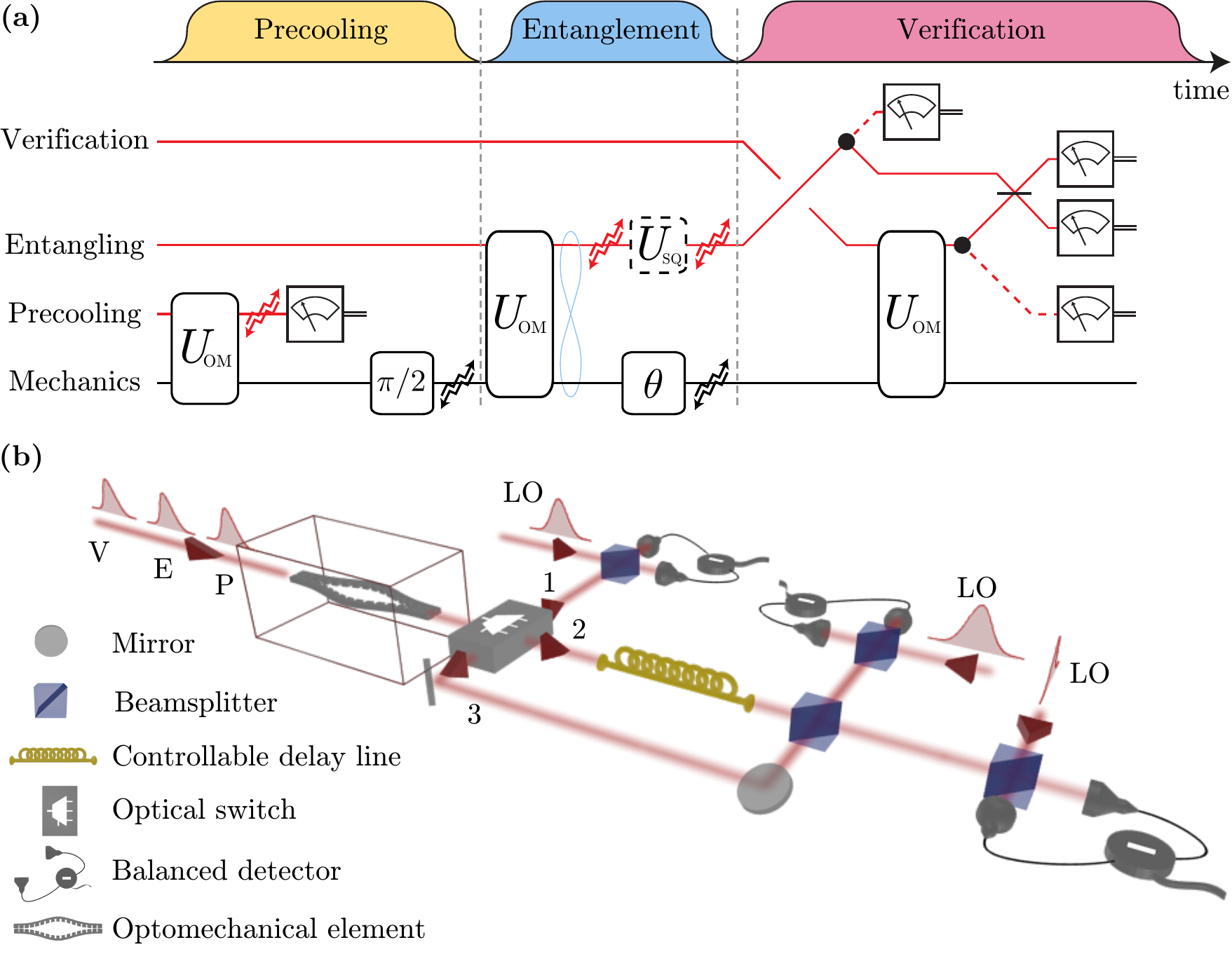}
\caption{{\textbf{Scheme for pulsed optical-mechanical entanglement preparation and verification.}} {\textbf{(a)}} Circuit diagram for the separate stages of the protocol---mechanical precooling, entanglement generation, and verification---with red lines corresponding to the optical modes and the black line representing the mechanical mode. The interaction between the optical pulses and the mechanical state is given by $U_{\textsc{om}} = \rme^{\rmi \chi \XL\XM}$. Following the precooling stage, the mechanical oscillator freely evolves over a phase-space angle $\theta=\pi/2$. Then, after the generation of the optical-mechanical entanglement we aim to verify, indicated by the blue figure eight, the mechanics rotates through a general angle $\theta$. Including an optional optical squeezer $U_{\textsc{sq}}$ increases the amount of optomechanical entanglement generated. The jagged arrows indicate optical loss or mechanical decoherence, while the meters represent optical homodyne measurements. Finally, the dashed red lines in the verification stage indicate the different paths the optical pulses may take. {\textbf{(b)}} The corresponding experimental design. For clarity we don't include the optional optical squeezer. Mechanical precooling is achieved by a direct phase-quadrature measurement of the precooling pulse (P) after the optomechanical interaction, and so the optical switch is set to position 1 during this stage. Verification of the optical and mechanical blocks of the optical-mechanical covariance matrix is achieved by directly homodyning the entangling (E) and verification pulses (V), respectively, and so the optical switch is again set to position 1. To access the correlations block, the entangling and verification pulses interfere on a beamsplitter and the optical output modes are homodyned. To realize this interference, the entangling pulse is sent down a delay line---with the optical switch in position 2---and then the verification pulse is sent towards the beamsplitter by moving the optical switch to position 3. LO refers to the coherent local oscillator pulse entering the homodyne detector.}
\label{singlesetup}
\end{figure}

The mechanical oscillator may be cooled prior to the generation of optical-mechanical entanglement by homodyning the output light following an optomechanical interaction. More specifically, this precooling stage is described by
\begin{equation}\label{omintmeas}
     \rho\rightarrow\mathcal{E}_{\gamma}({U}_{\textsc{rot}}(\pi/2)\ket{\PL}\bra{\PL}\circ\mathcal{E}_{\eta}({U}_{\textsc{om}}\circ\rho)).
\end{equation}
Mechanical precooling, Eq.~\eqref{omintmeas}, begins with a pulsed optomechanical interaction, followed by optical decoherence. Subsequently, a phase-quadrature measurement, with projector $\ket{\PL}\bra{\PL}$, is made on the light which projects the mechanical mode into a squeezed thermal state with a lower effective thermal occupation~\cite{Vanner2011}. The mechanical state is then allowed to evolve freely over a quarter of a period prior to the generation of optical-mechanical entanglement, in which time the mechanical mode undergoes some decoherence. In this way, Eq.~\eqref{omintmeas} prepares a mechanical state with reduced thermal noise along the momentum quadrature of phase space. The entanglement and precooling stages are illustrated in the circuit diagram of Fig.~\ref{singlesetup}(a).

%%%%%%%%%%%%%%%%%%%%%%%%%%%% Light-mechanics ent. %%%%%%%%%%%%%%%%%%%%%%%%%%%     
\section{Optical-mechanical entanglement}
\label{light-mechanics}
\subsection{Entangled state preparation and verification}
\label{Entangled state preparation and verification}
%Introduction and preparation
Our proposed experimental setup for the preparation of an entangled state of light and mechanics is shown in Fig.~\ref{singlesetup}(b). A pulse of light in a coherent state $\ket{\alpha}$ is injected into an optomechanical cavity, which then interacts via the entangling unitary $U_{\textsc{om}}$ with a mechanical oscillator in an initial thermal state, with mean thermal occupation $\bar{n}$. This interaction, followed by optical loss and mechanical decoherence is described by Eq.~\eqref{omint} and generates an entangled state of light and mechanics.

%Precooling pulse
Prior to the entangling stage, the precooling stage, described by Eq.~\eqref{omintmeas}, creates an initial mechanical state which can generate more entanglement for a given interaction strength $\chi$. This increase in performance is due to the reduction in thermal noise in the direction of the phase-space translations generated by the optomechanical unitary, and hence a higher amount of entanglement can be prepared between light and mechanics. Multiple precooling stages can be employed to further increase the degree of entanglement generated. However, within the range of parameters we consider in this work, we find that additional stages are unnecessary and lead to a negligible increase in the amount of entanglement generated. The analytic results of the precooling and entangling stages are given in \ref{symplectictransformations}.

%Verification description
Once an entangled state of light and mechanics has been generated, the full phase-space structure of the two-mode Gaussian state can be reconstructed. In this reconstruction process many runs of an experiment will be needed to build up adequate statistics for the different quadrature combinations. Experimentally, it will be important to track the first moments vector to accurately construct probability histograms of each observed quadrature. More specifically, in order to reconstruct the statistics of each quadrature the first moment must be recorded and subtracted from the measurement results, removing the effects of the random homodyne measurement outcomes from the data. Tracking the first moments vector may be achieved by considering the action of each Gaussian unitary and non-unitary process in the precooling and entangling stages of the protocol, as well as recording the measurement outcomes of the homodyne detection events. As mentioned in Section \ref{pulsed om}, when constructing the probability histograms of the mechanical quadratures, the projection of the deterministic momentum transfer along the mechanical quadrature under consideration may also be subtracted from the measurement results to avoid the necessity of a feedback force.

Turning our attention now to the reconstruction of the covariances, we refer to the reconstructed covariance matrix as $\sigma_{\textsc{ver}}$, which in block matrix form is
\begin{equation}\label{verCM}
\sigma_{\textsc{ver}}=\begin{pmatrix}A_{\textsc{ver}}&C_{\textsc{ver}}\\C_{\textsc{ver}}^{\mathrm{T}}&B_{\textsc{ver}}\end{pmatrix},
\end{equation}
and will approach $\sigma$ in the absence of decoherence processes. In our analysis, we assume that the interaction strength $\chi$ and the total optical efficiency $\eta$ are precisely known, which allows $\sigma_{\textsc{ver}}$ to be constructed accurately. So that the reader may easily follow the symplectic transformations, we present formulas for $\sigma_{\textsc{ver}}$ without the inclusion of optical loss in only the verification stage---optical losses are present in the precooling and entangling stages. However, including optical loss in the formulas for $\sigma_{\textsc{ver}}$ is straightforward and does not affect the results we present. In \ref{verfappendix}, we discuss how the elements of $\sigma_{\textsc{ver}}$ are obtained in more detail.

After the optomechanical interaction in the entangling stage of the protocol, the optical pulse exits the cavity and either a homodyne measurement is made on a rotated quadrature $\PL(\varphi)=\PL\cos\varphi-\XL\sin\varphi$, or the light is sent down a delay line. If the entangling pulse of light is directly homodyned, this allows the $A_{\textsc{ver}}$ block of $\sigma_{\textsc{ver}}$ to be determined and the required set of homodyne measurement angles may be noted from the expression
\begin{equation}\label{verA}
 \hspace*{-0.5cm}  A_{\textsc{ver}} =\begin{pmatrix}\mathrm{Var}(\XL)&\frac{1}{2}[\mathrm{Var}(\PL(\frac{3\pi}{4}))-\mathrm{Var}(\PL(\frac{\pi}{4}))]\\\frac{1}{2}[\mathrm{Var}(\PL(\frac{3\pi}{4}))-\mathrm{Var}(\PL(\frac{\pi}{4}))]&\mathrm{Var}(\PL)\end{pmatrix}.
\end{equation}

After the entangling pulse has been homodyned, a verification pulse in a pure coherent state is then introduced into the cavity after a controllable time delay $t=\theta/\omega_{\textsc{m}}$ in which time the mechanics evolves freely and undergoes some decoherence. This decoherence of the mechanical mode between the interaction with the entangling and verification pulses leads to imperfect reconstruction of the covariance matrix $\sigma$. After the optomechanical interaction, the quadrature vector of the verification pulse is $\mathbf{\Xv}(\theta)=(\Xv(\theta),\Pv(\theta))^{\mathrm{T}}$. The $\Xv(\theta)$ quadrature contains no information about the mechanical mode, however the phase quadrature of the verification pulse reads $\Pv(\theta)=P_{\textsc{in}}+\chi\XM(\theta)$, where the initial optical phase quadrature $P_{\textsc{in}}$ introduces noise in the reconstruction process and $\XM(\theta)=\XM\cos\theta+\PM\sin\theta$. Note that the argument $\theta$ of $\Pv(\theta)$ refers to the phase-space evolution of the mechanics and not a rotation in the optical subspace. To access the $B_{\textsc{ver}}$ block of the covariance matrix, optical homodyne measurements are made directly on the $\Pv(\theta)$ quadrature at various time delays and the input noise $\mathrm{Var}(P_{\textsc{in}})$ can be subtracted to increase the accuracy of the reconstruction process. Importantly, to verify the $B_{\textsc{ver}}$ block the homodyne measurement outcomes on the entangling pulse of light must be ignored so the light is traced out in the reconstruction process. The mechanical block is then given by
\begin{equation}\label{verB}
 \hspace*{-1cm}  B_{\textsc{ver}} =\frac{1}{\chi^2}\begin{pmatrix}\mathrm{Var}(\Pv(0))-\mathrm{Var}(P_{\textsc{in}})&\frac{1}{2}[\mathrm{Var}(\Pv(\frac{\pi}{4}))-\mathrm{Var}(\Pv(\frac{3\pi}{4}))]\\\frac{1}{2}[\mathrm{Var}(\Pv(\frac{\pi}{4}))-\mathrm{Var}(\Pv(\frac{3\pi}{4}))]&\mathrm{Var}(\Pv(\frac{\pi}{2}))-\mathrm{Var}(P_{\textsc{in}})\end{pmatrix}.
\end{equation}

To access the $C_{\textsc{ver}}$-block elements of the covariance matrix, the entangling light is not directly homodyned, but instead it is sent down a path towards a controllable delay line, while the verification pulse is sent down another path. These two pulses interact with each other on a 50:50 beamsplitter and each individual output is directly homodyned. Redirection of the entangling and verification pulses is achieved by using an optical switch as shown in Fig.~\ref{singlesetup}(b). The timescale over which this optical switch operates must be less than a quarter of the mechanical cycle as this corresponds to the time between the precooling pulse and the entangling pulse of light. This means there is no need for high-speed switching or small-scale integrated operation, one can use a larger switch that operates with very low optical loss, such as a Pockels cell. 

The beamsplitter operation, which mixes the entangling and verification pulses, is described by the $4\times4$ symplectic matrix $S_{\textsc{bs}}(\alpha,\beta)$ given in \ref{symplectictransformations}, where $\alpha$ and $\beta$ are the real beamsplitter parameters. To access every element of $C_{\textsc{ver}}$, two different beamsplitter configurations are required: $S_{\textsc{bs}}(\frac{\pi}{4},0)$ and $S_{\textsc{bs}}(\frac{\pi}{4},\frac{\pi}{2})$. These two configurations may be implemented with a single 50:50 beamsplitter and a controllable phase shifter. We define the quadrature vectors after the $S_{\textsc{bs}}(\frac{\pi}{4},0)$ and $S_{\textsc{bs}}(\frac{\pi}{4},\frac{\pi}{2})$ operations as $(\mathbf{\Xv'}(\theta),\mathbf{\XL'}(\theta))^{\mathrm{T}}$ and $(\mathbf{\Xv''}(\theta),\mathbf{\XL''}(\theta))^{\mathrm{T}}$, respectively. Then the $C_{\textsc{ver}}$ block are determined by
\begin{equation}\label{verC}
\hspace*{-0.75cm}   C_{\textsc{ver}} =
   \frac{1}{2\chi}
   \begin{pmatrix}\mathrm{Var}(\Pv''(2\pi))-\mathrm{Var}(\XL''(2\pi))&\mathrm{Var}(\Pv''(\frac{5\pi}{2}))-\mathrm{Var}(\XL''(\frac{5\pi}{2}))\\\mathrm{Var}(\Pv'(2\pi))-\mathrm{Var}(\PL'(2\pi))&\mathrm{Var}(\Pv'(\frac{5\pi}{2}))-\mathrm{Var}(\PL'(\frac{5\pi}{2}))\end{pmatrix}.
\end{equation}
Interference between the entangling and verification pulses is necessary to obtain the optomechanical covariances. Whereas, if there was no interference one would only be able to access the $A_{\textsc{ver}}$ and $B_{\textsc{ver}}$ blocks and not the full covariance matrix. Note, that we have included an addition of $2\pi$ in the arguments of the $ C_{\textsc{ver}}$ elements, which corresponds to a full rotation of the mechanical mode in phase space. The necessity of this additional $2\pi$ is explained in Section \ref{nonmoncons}. Furthermore, when the additional factor of $2\pi$ is included, the time between the entangling and verification pulses is always greater than the time between the precooling and entangling pulses.

%Entanglement
In order to quantify entanglement, we employ the logarithmic negativity, which is a well-suited entanglement monotone for low-dimensional discrete variable systems~\cite{peres1996separability,HORODECKI19961,lee2000partial,PhysRevLett.90.027901} and for bipartite Gaussian states~\cite{simon2000peres,vidal2002computable,PhysRevLett.86.3658}. This monotone quantifies the extent to which the positivity of the density operator is violated after a transpose operation is applied to a single mode. For bipartite Gaussian systems, partial transposition leads to a covariance matrix with eigenvalues given by
\begin{align}
    \tilde{\nu}_{\pm} = \sqrt{\frac{\tilde{\Delta} \pm \sqrt{\tilde{\Delta}^2 - 4 \det \sigma}}{2}} 
\end{align}
where $\tilde{\Delta} = \det A + \det B - 2 \det C$. A necessary and sufficient condition for bipartite entanglement in Gaussian continuous variable systems is given by $\tilde{\nu}_{-}<1/2$. The logarithmic negativity of the two-mode state with covariance matrix $\sigma$ is then calculated as 
\begin{equation}
 E_{\mathcal{N}}(\sigma) = \max\{0, -\log_2 (2\tilde{\nu}_{-})\}. 
\end{equation}

\subsection{Optical squeezing and conservative approaches to entanglement estimation}\label{nonmoncons}

Having now outlined the protocol for optical-mechanical entanglement preparation and verification, we now consider two additional subtleties the protocol presents. Firstly, we investigate the non-monotonic behaviour in the logarithmic negativity with increasing interaction strength in the presence of both optical loss and mechanical decoherence, and how the protocol can be made more resilient to these decoherence effects by using optical squeezers. Secondly, we consider the time order in which elements of the covariance matrix must be measured to conservatively estimate the entanglement generated in the protocol. To facilitate this quantitative discussion, in Table \ref{parameters} we list values for our system parameters based on the sliced-silicon nanobeam architecture comprising the optomechanical system that appears in Refs.~\cite{muhonen2019state,leijssen2015strong,leijssen2017nonlinear}. In the following, we also discuss the parameters we change from these experiments, namely the thermal occupation $\bar{n}$ and the optical efficiency $\eta$. 

In Ref.~\cite{muhonen2019state}, a reduction in the thermal occupation from $\bar{n}=22,200$ to an effective occupation of $3,400$ was achieved using a series of pulsed-optomechanical interactions at $3.2~\mathrm{K}$. Therefore, multiple precooling steps can be used to increase the temperature at which the protocol can operate at. However, to simplify the discussions, we do not consider multiple precooling steps and we assume a bath temperature of approximately $0.1~\mathrm{K}$, hence $\bar{n}=\bar{N}=500$. Such temperatures are readily achieved with commercially available dilution refrigerators. Furthermore, in the table we propose an improved optical efficiency of $\eta=0.855$. Optical efficiencies of $1.3\%$ have been achieved in experiments, and without implementing further cooling techniques a reduction in the uncertainty below the zero-point fluctuations may be achieved with $\eta>8\%$~\cite{muhonen2019state,leijssen2017nonlinear}. In Section \ref{comparisonsection}, we further discuss the requirements on optical efficiency that our protocols present. In particular, the optical-mechanical entanglement protocol can generate entanglement at very low optical efficiencies, less than $1\%$, while the protocols for generating mechanical entanglement require total optical efficiencies greater than $10\%$. As noted in Ref.~\cite{leijssen2017nonlinear}, improvements to the optical design and employing new waveguide coupling techniques provide an encouraging route for further experimental progress in pulsed optomechanics.

\begin{table}[h]
 \caption{\textbf{List of proposed experimental parameters for the optical-mechanical and mechanical entanglement protocols.} The parameter values are based on the sliced-silicon nanobeam system that appears in Refs~\cite{muhonen2019state,leijssen2015strong,leijssen2017nonlinear}. Here, we assume that the mechanical oscillator and its environment are in equilibrium, hence $\bar{n}=\bar{N}=500$---this corresponds to a temperature of approximately $0.1~\mathrm{K}$. We propose an improved optical efficiency of $\eta=0.855$. This optical efficiency may be decomposed into a part that accounts for output cavity coupling losses from the cavity, $\eta_{\textsc{cav}}$, and a part that accounts for homodyne detection efficiency $\eta_{\textsc{det}}$. In this work, we consider a range of values for the optomechanical interaction strength $\chi$.
}
 %%%%%
\centering
\begin{tabular}{c|c}
\hline\hline
System parameter & Value \\
\hline
   $ \omega_{\textsc{m}}/2\pi$ & $4~\mathrm{MHz}$  \\
     $\gamma/2\pi$ & $100~\mathrm{Hz}$\\
      $\bar{n}=\bar{N}$ & $500$ \\
      $\kappa/2\pi$ & $20~\mathrm{GHz}$\\
     $g_{0}/2\pi$ & $30~\mathrm{MHz}$ \\
     %\chi & [0,100] \\
     $\eta_{\textsc{cav}}$ & $0.9$  \\
     $\eta_{\textsc{det}}$ & $0.95$  \\
     %\eta_{\mathrm{tot}} & 0.855  \\
     \hline\hline
     \end{tabular}
     \label{parameters}
     \end{table}

%Non-monotonic behaviour
After the optomechanical interaction between the entangling pulse and the precooled mechanical oscillator, the reduced state of the optical mode will satisfy $\mathrm{Var}(\PL)>\mathrm{Var}(\XL)$. This is due to the phase-space displacements generated by $U_{\textsc{om}}$ along the $\PL$ axis of optical phase space, which are proportional to the mechanical position quadrature operator. The optical mode is then subject to the phase-insensitive decoherence channel $\mathcal{E}_{\eta}$, which introduces vacuum noise and degrades the optomechanical correlations. However, to reduce these deleterious effects, an optical squeezer may be applied immediately after the optomechanical interaction. Optical squeezing is a useful way to protect single-mode quantum states from interactions with the environment, for example squeezers may be used to improve state fidelity in teleportation~\cite{Mista2005teleportation} and the performance of quantum memory devices~\cite{Filip2008Excess}. Furthermore, the utility of such operations in protecting quantum coherence has been investigated theoretically~\cite{serafini2004minimum,Filip2013Gaussian,brewster2018reduced} and verified experimentally~\cite{le2018slowing}. Unlike these recent applications of optical squeezing, in this work we utilize optical squeezing to protect the entanglement of a bipartite state specifically. Eq.~\eqref{omint}, which describes the generation of optical-mechanical entanglement, is therefore modified to
\begin{equation}\label{omintsq}
     \rho\rightarrow\mathcal{E}_{{\gamma}}(U_{\textsc{rot}}(\theta)\circ\mathcal{E}_{{\eta}_{\textsc{det}}}(U_{\textsc{sq}}\circ\mathcal{E}_{{\eta_{\textsc{cav}}}}({U}_{\textsc{om}}\circ\rho))).
\end{equation}
Here, the squeezing unitary is $U_{\textsc{sq}}=\exp{\frac{r}{2}[({a}^{\dagger})^2-{a}^2]}$, with real squeezing parameter $r$. Moreover, the channel $\mathcal{E}_{\eta_{\textsc{cav}}}$ represents losses from the optical cavity and $\mathcal{E}_{\eta_{\textsc{det}}}$ is the channel describing detection inefficiencies. We assume that the optical squeezer can be turned off during the precooling and verification stages of the protocol. However, if for a particular experimental implementation of our scheme, operating the squeezer in this way is not feasible, then the description of the precooling and verification stages is straightforwardly modified by using the symplectic transformation $S_{\textsc{sq}}(r)$, which appears in \ref{symplectictransformations}. 

In our entanglement scheme, we are interested in how squeezing operations allow for the protection of entanglement---rather than quantum superposition as studied in Refs.~\cite{serafini2004minimum,Filip2013Gaussian,brewster2018reduced,le2018slowing}.  For a Gaussian bosonic channel, it is well known that the output state with maximum purity and minimum von-Neumann entropy corresponds to a coherent state input~\cite{Giovannetti2004,mari2014quantum}. Likewise, when a squeezing operation is applied to an input state of fixed given purity, the input state that maximizes the output purity and minimizes the output entropy, with respect to the real squeezing parameter $r$, is given by a symmetric mode in phase space, i.e. a thermal state, see \ref{channel}. This may suggest that the optical squeezing parameter $r_{\textsc{sym}}=\frac{1}{4}\ln\left[1-\eta_{\textsc{cav}}+\eta_{\textsc{cav}}(1+2V_{\textsc{x}}\chi^2)\right]$, which symmetrizes the optical mode such that $\mathrm{Var}(\PL)=\mathrm{Var}(\XL)$, is a good candidate for the optimal squeezing parameter to also maximize the output logarithmic negativity. Here, $V_{\textsc{x}}$ is the position variance of the precooled mechanical state, which is derived in \ref{symplectictransformations}.

However, the entangling unitary $U_{\textsc{om}}$  generates optomechanical correlations which are distributed asymmetrically between the optical phase and amplitude quadratures. Namely, only the optical phase quadrature contains information about the mechanical position, which may be lost to the optical environment leading to a reduction in the logarithmic negativity. Furthermore, the optical environment will be less sensitive to displacements along the phase quadrature for an optical state of a given purity with $\mathrm{Var}(\PL)>\mathrm{Var}(\XL)$. Therefore, a state with $\mathrm{Var}(\PL)>\mathrm{Var}(\XL)$ at the input of the optical loss channel $\mathcal{E}_{\eta_{\textsc{det}}}$, reduces the sensitivity of the optical environment to the mechanical position quadrature and the optomechanical correlations. 

\begin{figure}
\vspace{-3cm}
\includegraphics[width=0.89167\linewidth,angle=270]{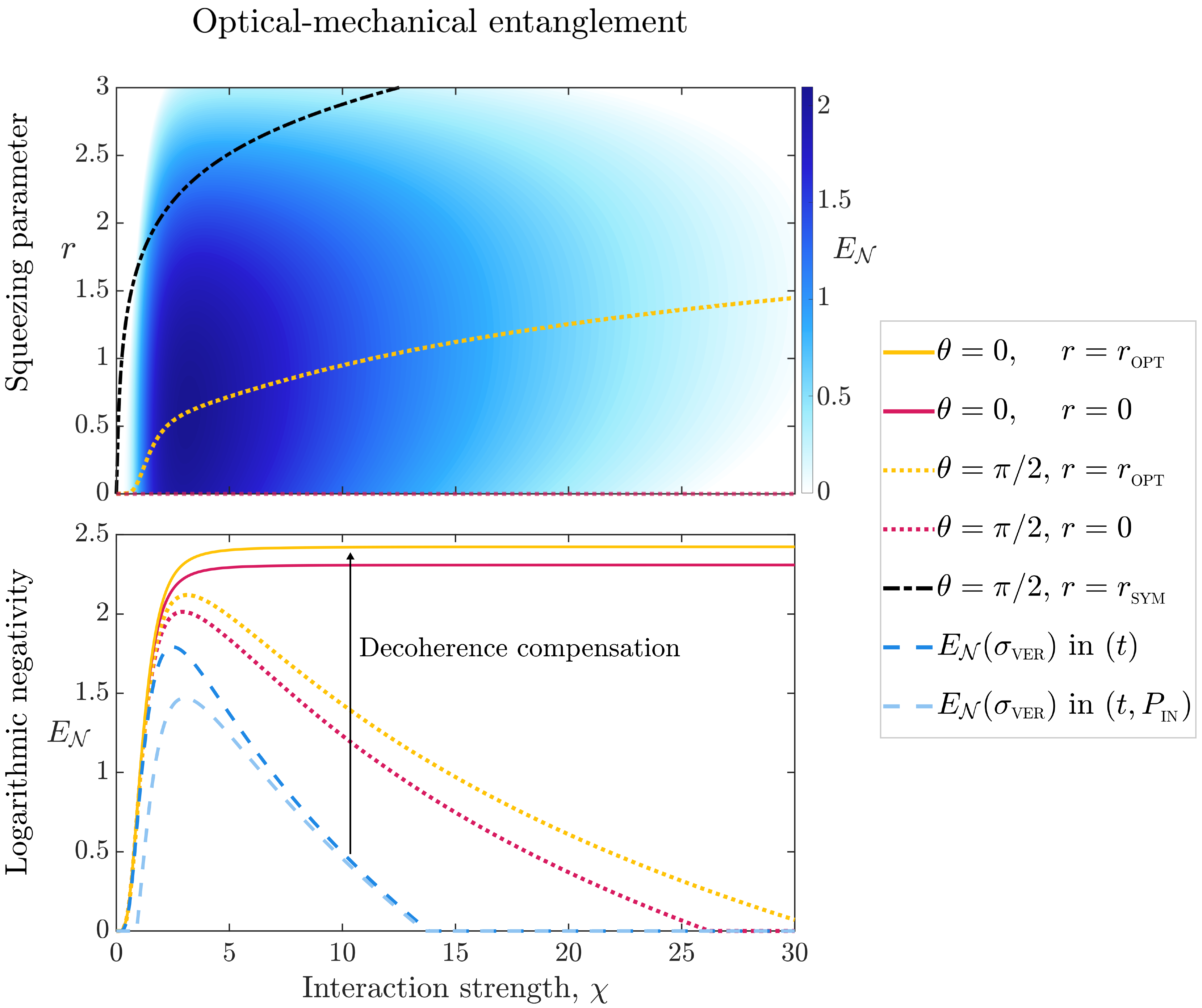}
\caption{{\textbf{Logarithmic negativity $E_{\mathcal{N}}(\sigma)$ generated between a pulse of light and a precooled mechanical state.}} {\textbf{Upper:}} Contour plot of logarithmic negativity $E_{\mathcal{N}}(\sigma)$ as a function of optomechanical interaction strength $\chi$ and the optical squeezing parameter $r$---with other system parameters corresponding to those listed in Table \ref{parameters}. Here, the mechanical state has evolved, and decohered, through a phase-space angle of $\theta=\pi/2$ after the time of entanglement generation. The maximum achievable logarithmic negativity is $E_{\mathcal{N}}=2.12$, which requires $\chi=3.12$ and $r=0.60$. Without the use of an optical squeezer, a value of $E_{\mathcal{N}}=2.01$ is obtained with $\chi=2.94$. The squeezing parameter $r_{\textsc{sym}}$ symmetrizes the optical mode after the optomechanical interaction, while $r=r_{\textsc{opt}}$ maximizes $E_{\mathcal{N}}(\sigma)$ as a function of $\chi$. The contour plot shows $r_{\textsc{sym}}>r_{\textsc{opt}}$ for all $\chi$, which demonstrates that an asymmetric optical phase-space distribution is optimal for maximizing entanglement.
 {\textbf{Lower:}} Plot showing logarithmic negativity as a function of interaction strength $\chi$ in the presence of both optical loss and mechanical decoherence. The plot also shows that two conservative approaches, which ensure that the logarithmic negativity is not overestimated, can verify a siginificant amount of entanglement, i.e. $E_{\mathcal{N}}(\sigma_{\textsc{ver}})>1$, for a range of $\chi$. Here, $\theta$ refers to the angle through which the mechanical oscillator evolves in phase-space after the optomechanical interaction. The conservative approach in time, which accounts for different times between the entangling and verification pulses, is denoted by $(t)$, while the conservative approach in time and noise, which also accounts for lack of knowledge about the optical noise on the verification pulse, is denoted by $(t,P_{\textsc{in}})$. At large $\chi$ the two conservative approaches show the same limiting behaviour as the contribution of the optical noise becomes less relevant. The arrow indicates that better approximations of $\sigma$ can be made from $\sigma_{\textsc{ver}}$ by applying the inverse of the mechanical decoherence map to each element as noted in the main text and explored further in \ref{verfappendix}.}
    \label{logneg1}
\end{figure}

A balance between a symmetric optical phase-space distribution, which minimises the output entropy of the optical decoherence channel, and an optical phase-space distribution with $\mathrm{Var}(\PL)>\mathrm{Var}(\XL)$, which reduces the sensitivity of the optical environment to optomechanical correlations, leads to the optimal squeezing parameter $r_{\textsc{opt}}$ that maximizes the logarithmic negativity. In the upper plot of Fig.~\ref{logneg1}, $r_{\textsc{opt}}$ is plotted as a function of the interaction strength $\chi$. This plot shows the logarithmic negativity of the optical-mechanical entangled state as a function of $r$ and $\chi$. Here, the entangled state has been subject to both optical and mechanical decoherence processes since the time of entanglement generation. For comparison, we also present $r_{\textsc{sym}}$ as a function of $\chi$, and we note that $r_{\textsc{sym}}>r_{\textsc{opt}}$ for all $\chi$---demonstrating the balance between the two effects described above. The experimental implementation of this squeezing operation, which we introduce to protect entanglement in pulsed optomechanics, is made feasible by experimental advances in squeezing pulses of light~\cite{Slusher1987pulsed,Ourjoumtsev2006generating} and non-classical continuous travelling waves of light~\cite{Miwa2014exploring}.

Importantly, as is evident from the lower plot of Fig.~\ref{logneg1}, non-monotonic behaviour in the logarithmic negativity is observed when both optical loss and mechanical decoherence are present after the generation of entanglement. The existence of this non-monotonic behaviour is because---at large values of $\chi$---the increase in the magnitude of the optomechanical correlations with interaction strength scales in the same way as the decrease in the logarithmic negativity due to the decoherence of either the optical or mechanical mode alone. Hence, when both optical and mechanical decoherence processes are present, the logarithmic negativity first rises to a maximum before tending towards zero as $\chi$ is increased and decoherence processes outweigh the effect of the entangling optomechanical unitary. When the optical squeezing operation can be applied before the decoherence channels, the non-monotonic behaviour in logarithmic negativity can be completely eliminated. In this case, the entangled state is made more resilient to loss of information to the environment before it enters the decoherence channels and so higher values of $E_{\mathcal{N}}(\sigma)$ can be reached than in Fig.~\ref{logneg1}.  We refer the reader to \ref{nonmonappendix} for a discussion of this case.

% Detail the two conservative approaches --- conservative in time and wrt noise.
The lower plot in Fig.~\ref{logneg1} also demonstrates how accurately the optical-mechanical entanglement is measured in the verification process. As it is necessary to allow the mechanical oscillator to undergo free mechanical evolution to obtain information about the rotated mechanical quadratures $\XM(\theta)$, we encounter the problem that different elements of the covariance matrix must be measured at different times, having suffered different amounts of mechanical decoherence. Moreover, care must be taken when calculating the logarithmic negativity from the reconstructed covariance matrix, in particular one must ensure that  $E_{\mathcal{N}}(\sigma)\geq{E}_{\mathcal{N}}(\sigma_{\textsc{ver}})$, meaning that the logarithmic negativity is not overestimated. To ensure this condition, we demand that the elements of $C_{\textsc{ver}}$ are accessed at later times than those of $A_{\textsc{ver}}$ and $B_{\textsc{ver}}$, hence the addition of $2\pi$ in the argument of the elements in Eq.~\eqref{verC}. Therefore, the $C_{\textsc{ver}}$ elements experience a larger amount of decoherence than the $A_{\textsc{ver}}$ and $B_{\textsc{ver}}$ elements, meaning that the optical-mechanical correlations cannot be falsely enhanced in the verification process. We refer to this approach as \textit{conservative in time}, and the conservative nature of this approach is demonstrated in the lower plot of Fig.~\ref{logneg1}. Moreover, we consider the case in which state reconstruction is performed without assuming that the noise statistics of the verification pulse $\mathrm{Var}({P_{\textsc{in}}})$ is known and can be subtracted. Similarly, we refer to this approach as \textit{conservative in time and noise}. 

If the mechanical damping rate $\gamma$ and thermal occupation $\bar{n}$ are well known, the inverse of the mechanical decoherence may be applied to each element of $\sigma_{\textsc{ver}}$ to compensate for decoherence, as illustrated in the lower plot of Fig.~\ref{logneg1} by the vertical arrow. The application of the inverse map to the elements of $\sigma_{\textsc{ver}}$ that depend on two different times is more involved and is described in \ref{verfappendix}. Also, if decoherence can be accurately compensated for in this way, the need to be conservative with respect to time is no longer needed and so additional factors of $2\pi$ can be removed from the arguments of the $C_{\textsc{ver}}$-elements.

%%%%%%%%%%%%%%%%%%%%%%%%%%%%%%%%%%%

\section{Mechanical entanglement}\label{twomechanics}
Our proposal for preparing an entangled state of two mechanical oscillators builds upon the optical-mechanical entanglement described in the previous section. In this section, we describe schemes for converting entanglement between two optical-mechanical systems to entanglement between two mechanical oscillators using optical homodyne measurements. In particular, we introduce two such complementary schemes. One scheme employs an optical interferometric design and therefore necessitates the introduction of two optical modes, while the other scheme comprises a non-interferometric setup. For each scheme, we compute the amount of entanglement that can be prepared and verified.

\subsection{Interferometric scheme}\label{twomechanicspart1}
\begin{figure}
\centering
\includegraphics[width=1.15\linewidth]{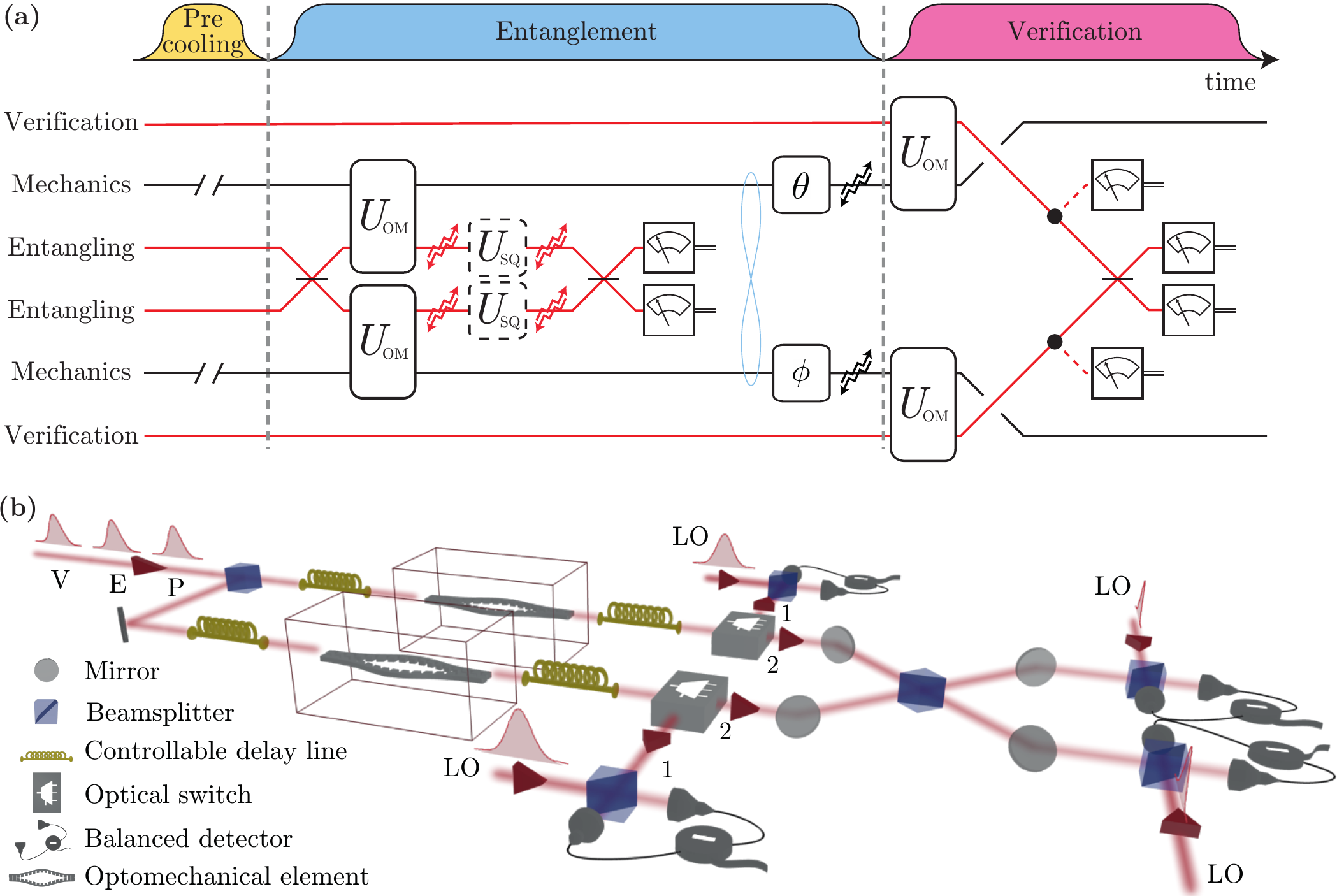}
\caption{\textbf{Interferometric scheme for mechanical entanglement preparation and verification.} {\textbf{(a)}} Circuit diagram for the preparation and verification stages after the precooling stage has taken place. The angles through which the first and second mechanical modes freely evolve are given by $\theta$ and $\phi$, respectively. {\textbf{(b)}} Proposed experimental setup to implement the protocol. By virtue of the optical switches and controllable delay lines, the setup allows for mechanical precooling, entangled state preparation, and verification. Each mechanical mode is initially precooled by an interaction with a precooling pulse, followed by an optical phase-quadrature measurement. This is achieved by setting the optical switches to position 1. Mechanical entanglement is generated by first establishing pairwise optical-mechanical entanglement, and then converting this to mechanical entanglement via optical interference and homodyne measurements. Setting the optical switches to position 2 allows for this. The blocks of the covariance matrix which correspond to each mechanical oscillator are reconstructed by interacting each mechanical mode with a verification pulse and directly homodyning the optical outputs, which requires the optical switches to be set to position 1. Instead, if the optical switches are set to position 2, then the mechanical-mechanical correlations can be measured.
}
\label{doublemechsetup}
\end{figure}

%Outline of protocol
The interferometric protocol to entangle two mechanical modes is shown in the circuit diagram of Fig.~\ref{doublemechsetup}, where, for brevity, the details of the precooling stage have been omitted. The entanglement stage of the protocol starts with a pulse of light in a coherent state impinging on one input port of a Mach-Zehnder interferometer. The other input port is in the vacuum state, such that after the first 50:50 beamsplitter interaction, described by $U_{\textsc{bs}}$, two coherent pulses of light propagate in the upper and lower arms towards the two mechanical modes. We label the quadratures of the optical modes $\mathbf{X}_{\textsc{l}\scriptscriptstyle{1}}=(\XLo,\PLo)^{\textsc{T}}$ and $\mathbf{X}_{\textsc{l}\scriptscriptstyle{2}}=(\XLt,\PLt)^{\textsc{T}}$, and the mechanical modes with which these light modes interact as $\mathbf{X}_{\textsc{m}\scriptscriptstyle{1}}$ and $\mathbf{X}_{\textsc{m}\scriptscriptstyle{2}}$, respectively. Each light mode interacts with its respective mechanical mode through the unitary $U_{\textsc{om}}$, hence the total optomechanical interaction is described by $U_{\textsc{om}}\otimes{U}_{\textsc{om}}=U_{\textsc{om}}^{\otimes2}$. After these interactions, the decoherence channel $\mathcal{E}_{{\eta_{\textsc{cav}}}}$ describing cavity losses acts on the two-mode optical subspace, and optional unitary squeezing operations $U_{\textsc{sq}}^{\otimes2}$ may be applied to each optical mode to increase the final amount of entanglement generated. Following this, the optical modes experience the loss channel $\mathcal{E}_{{\eta_{\textsc{det}}}}$ and then pass through a 50:50 beamsplitter, which erases which-path information about the optomechanical interactions. In our model, the ordering of the final loss channel and beamsplitter is arbitrary as they both lead to the same final mechanical state. To turn the optical-mechanical entanglement to mechanical entanglement, optical homodyne measurements are made on the rotated quadratures $\XLo(\varphi)$ and $\XLt(\psi)$, described by the projector $\ket{\XLo(\varphi),\XLt(\psi)}\bra{\XLo(\varphi),\XLt(\psi)}$. After the mechanical entanglement has been generated, the first and second mechanical modes evolve through phase-space angles $\theta$ and $\phi$, respectively, which is described by $U_{\textsc{rot}}(\theta,\phi)=U_{\textsc{rot}}(\theta)\otimes U_{\textsc{rot}}(\phi)$. Different periods of mechanical free evolution can be achieved using controllable delay lines, as is shown in Fig.~\ref{doublemechsetup}(b) and discussed in further detail below. During this free evolution the mechanical oscillators decohere via the channel $\mathcal{E}_{{\gamma}}$.
Including mechanical decoherence, the interferometric entanglement protocol is therefore given by
\begin{gather}\label{DMMap}
\rho\rightarrow\mathcal{E}_{\gamma}(U_{\textsc{rot}}(\theta,\phi)\ket{\XLo(\varphi),\XLt(\psi)}\bra{\XLo(\varphi),\XLt(\psi)} U_{\textsc{bs}}\circ\\\nonumber
\mathcal{E}_{\eta_{\textsc{det}}}( U_{\textsc{sq}}^{\otimes2}\circ\mathcal{E}_{\eta_{\textsc{cav}}}(U_{\textsc{om}}^{\otimes2}U_{\textsc{bs}}\circ\rho))).
\end{gather}

%%%%Entanglement
From Eq.~\eqref{DMMap}, one finds that at the input ports of the homodyne detectors the optical phase quadratures $\PLo$ and $\PLt$ contain information about the mechanical EPR quadratures $\XMo\pm\XMt$, whilst no knowledge of mechanical position can be gleaned from the the corresponding optical amplitude quadratures, $\XLo$ and $\XLt$. In what follows, we therefore choose that the homodyne angles are set to one of two equivalent configurations given by $(\varphi,\psi)=\left\{(0,\pi/2),(\pi/2,0)\right\}$. These configurations correspond to Bayesian inference of a mechanical EPR quadrature, which projects the mechanics into an entangled state. Conversely, if both phase quadratures are measured, then no mechanical entanglement can be created, as this would allow information about each mechanical quadrature to be obtained separately. Interestingly, there exists other optimal homodyne angles than those stated above, which also maximize the logarithmic negativity. These angles correspond to a balance between the Bayesian inference of an EPR quadrature and an effective unitary operation, and this is discussed further in \ref{MeasOpApproach}.

% Verification
The state prepared by the precooling and entangling stages, as illustrated in Fig.~\ref{doublemechsetup}, can be reconstructed by sending a subsequent verification pulse into the input port of the interferometer. The verification pulses in the upper and lower arms then interact with the mechanical oscillators after time delays $t_{1}=\theta/\omega_{\textsc{m}}$ and $t_{2}=\phi/\omega_{\textsc{m}}$, respectively. The phase quadrature of the verification pulse in the upper arm of the Mach-Zehnder interferometer evolves from $P_{\textsc{in}\scriptscriptstyle{1}}$ to $\PVo(\theta) = P_{\textsc{in}\scriptscriptstyle{1}} + \chi \XMo(\theta)$, during the optomechanical interaction. While the verification pulse in the lower arm evolves from $P_{\textsc{in}\scriptscriptstyle{2}}$ to $\PVt(\phi)=P_{\textsc{in}\scriptscriptstyle{2}} + \chi \XMt(\phi)$. As in the optical-mechanical entanglement scheme, for clarity, we present results for the verification procedure without the inclusion of optical loss and refer the reader to \ref{verfappendix} for a more complete discussion, including optical loss.

\begin{figure}
\vspace{-3cm}
\includegraphics[width=0.89167\linewidth,angle=270]{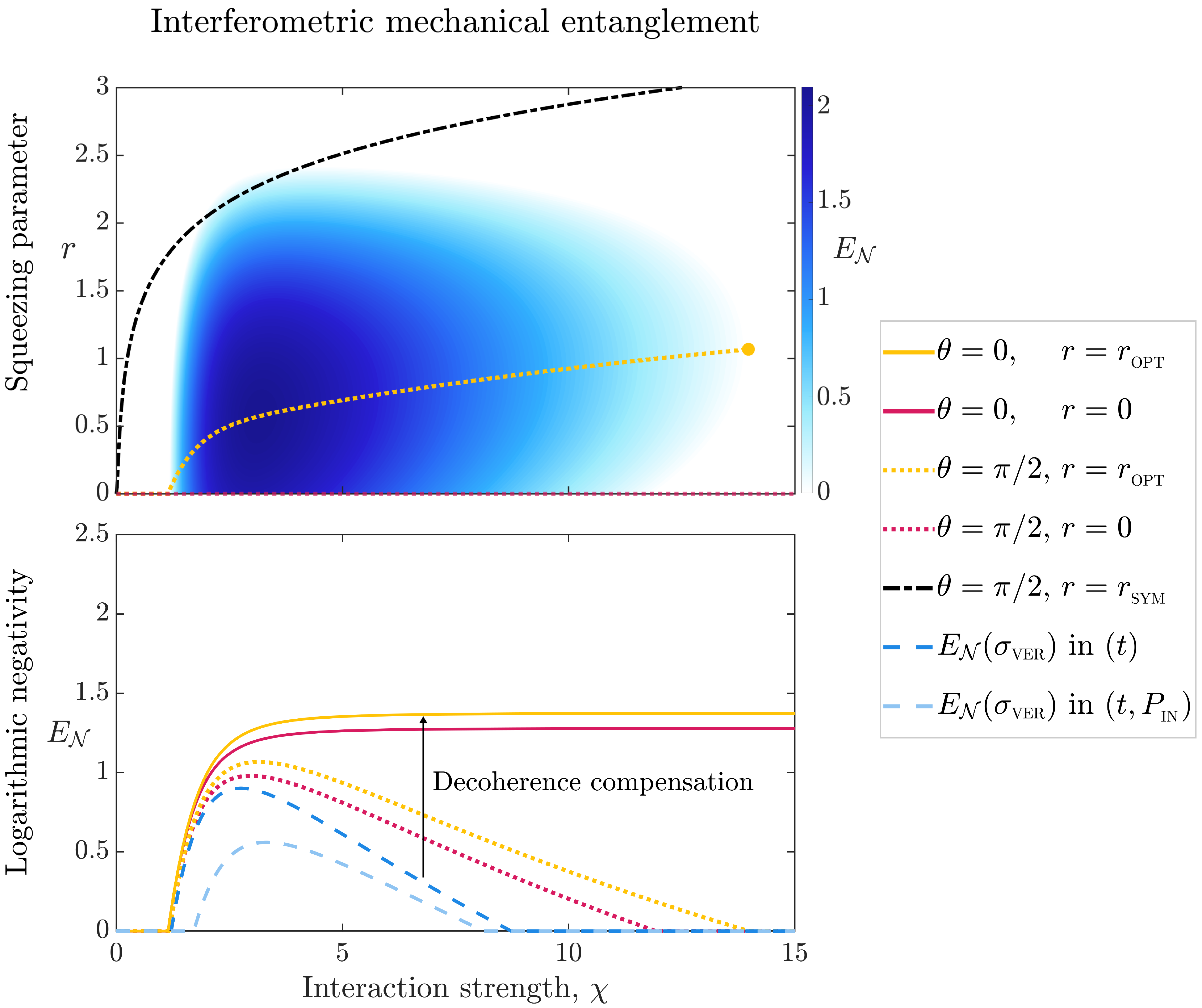}
\caption{ \textbf{Logarithmic negativity $E_{\mathcal{N}}(\sigma)$ of the entangled mechanical state generated by our interferometric scheme.} \textbf{Upper:} Contour plot of logarithmic negativity $E_{\mathcal{N}}(\sigma)$ as a function of optomechanical interaction strength $\chi$ and squeezing parameter $r$. Here, the other system parameters correspond to those listed in Table \ref{parameters}. Since the time of entanglement generation, each mechanical oscillator has rotated, and decohered, over a quarter of a mechanical period. The maximum logarithmic negativity is $E_{\mathcal{N}}=1.07$, which corresponds to $\chi= 3.15$ and $r=0.57$. Without the use of a squeezer the protocol achieves $E_{\mathcal{N}}=0.98$ with $\chi=2.97$. The curve corresponding to the optimal squeezing parameter terminates at the point where, even with squeezing, no entanglement can be generated. We note that the formula for $r_{\textsc{sym}}$ is the same as in the optical-mechanical case. \textbf{Lower:} Logarithmic negativity in the presence of both optical loss and mechanical decoherence after state preparation, and conservative approaches for entanglement verification. The plot shows the case where the two mechanical modes have freely evolved through the same phase-space angles, $\theta=\phi$, and hence have decohered by the same amount. Here, $r_{\textsc{opt}}$ is the optimal squeezing parameter to maximize the logarithmic negativity. The arrow indicates that with an inverse transform, better approximations to the entanglement generated at $\theta=\phi=0$ can be made from the statistics gathered using the conservative approaches.}
\label{intlogneg}
\end{figure}

We represent the covariance matrix obtained in the verification procedure as $\sigma_{\textsc{ver}}$, which has the same block-matrix form as Eq.~\eqref{verCM}. However, now both the $A_{\textsc{ver}}$ and $B_{\textsc{ver}}$ blocks correspond to a covariance matrix of a reduced mechanical state. These matrix blocks are obtained by performing phase homodyne measurements on the verification pulses after they interact with the mechanical modes. To allow for this direct detection, optical switches are placed after the optomechanical interactions in Fig.~\ref{doublemechsetup}(b) to redirect the verification pulses away from the second 50:50 beamsplitter of the Mach-Zehnder interferometer. The elements of the $A_{\textsc{ver}}$ and $B_{\textsc{ver}}$ blocks, which depend on the mechanical phase-space angles $\theta$ and $\phi$, respectively, are therefore given by equivalent expressions to Eq.~\eqref{verB}---with $\Pv(\theta)$ replaced by $\PVo(\theta)$ or $\PVt(\phi)$.
    
On the other hand, the correlations between the two mechanical modes, defined by the $C_{\textsc{ver}}$-block elements, are obtained by mixing the two verification pulses on a beamsplitter after the optomechanical interactions. Similarly to the verification of optical-mechanical entanglement, it is the interference of these two optical pulses that allows for the full reconstruction of the covariance matrix. After this operation, the resulting phase quadrature operators at the beamsplitter outputs are given by $P_{\textsc{v}\scriptscriptstyle{\pm}}(\theta,\phi) = (\PVo(\theta) \pm \PVt(\phi))/\sqrt{2}$. Homodyne measurements are then performed on these quadratures in order to construct $C_{\textsc{ver}}$. Introducing $\delta(\theta,\phi)=	\mathrm{Var}(P_{\textsc{v}\scriptscriptstyle{+}}(\theta, \phi)) - \mathrm{Var}(P_{\textsc{v}\scriptscriptstyle{-}}(\theta, \phi))$ allows one to write $C_{\textsc{ver}}$ as
\begin{eqnarray} \label{CBlockDM}
C_{\textsc{ver}} = \frac{1}{2\chi^2} \begin{pmatrix}
	 \delta(2\pi,2\pi)&   \delta(2\pi,\frac{5\pi}{2}) \\
	\delta(\frac{5\pi}{2},2\pi) & \delta(\frac{5\pi}{2},\frac{5\pi}{2})\end{pmatrix}.
\end{eqnarray}
The off-diagonal elements of $C_{\textsc{ver}}$, correspond to different periods of free evolution for each mechanical mode. To introduce this difference---and to ensure the recombination of the verification pulses on the final beamsplitter---controllable delay lines may be inserted before and after the optomechanical interactions in the interferometer of Fig.~\ref{doublemechsetup}(b). We have again included an additional factor of $2\pi$ in the arguments of the $C_{\textsc{ver}}$-block such that the logarithmic negativity is conservatively estimated as discussed in Section \ref{nonmoncons}.

Fig.~\ref{intlogneg} shows the logarithmic negativity of the entangled state of two mechanical oscillators, and demonstrates how optical squeezing may be used to increase the amount of entanglement generated. The squeezing changes the phase-space distributions of the light modes to ensure the protocol remains robust to decoherence processes in the large $\chi$ limit. The reduced state of each mechanical mode is identical---see \ref{symplectictransformations}---meaning the Gaussian state is symmetric, which allows the entanglement of formation to be computed analytically~\cite{Bennett1996,Giedke2003}. But as this entanglement measure is also a monotonically decreasing function of $\tilde{\nu}_{-}$, it is an equivalent measure to $E_{\mathcal{N}}$. The lower plot of Fig.~\ref{intlogneg} also reaffirms the conclusions of Section \ref{nonmoncons}. Namely, that the origin of the non-monotonic behaviour in the logarithmic negativity is due to the presence of both optical loss and mechanical decoherence. Conservative approaches for state verification are also shown in the lower plot of Fig.~\ref{intlogneg}. As in the case of the optical-mechanical entanglement protocol, the statistics obtained via the conservative approaches may be compensated for by using an inverse decoherence map.

The interferometric scheme we have introduced  above offers an alternative strategy to the mechanical entanglement scheme presented in Ref.~\cite{pirandola2006macroscopic}---which utilizes Stokes scattering in the resolved-sideband regime. Namely, in this work, the potential to generate entanglement using short optical pulses in the unresolved sideband regime is explored, which allows rapid preparation and verification of entanglement, and we introduce and detail a full-state-characterization procedure using these short optical pulses.

\subsection{Non-interferometric scheme}
\begin{figure}
\includegraphics[width=1.15\linewidth]{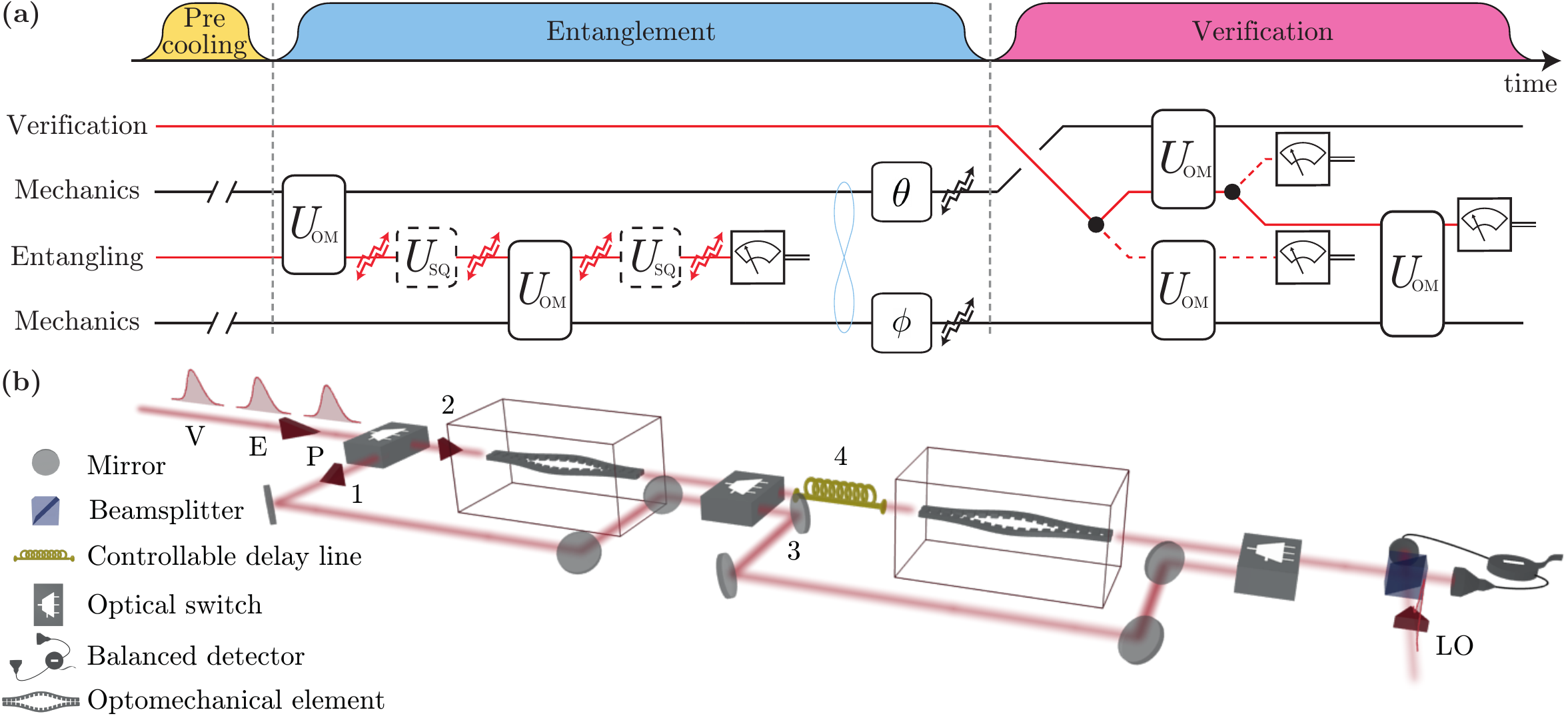}
\caption{\textbf{Non-interferometric scheme for mechanical entanglement preparation and verification.} {\textbf{(a)}} Circuit diagram of the protocol. {\textbf{(b)}} Proposed experimental setup to implement the scheme. Here, a single homodyne detector and controllable delay line suffice to realize the precooling, entangling, and verification stages of the protocol.
%precooling
For each mechanical mode, the precooling stage consists of an interaction with the precooling pulse followed by a phase-quadrature measurement on the output pulse. For the first mechanical mode, this is achieved using the switch settings 2 and 3, respectively, while for the second mechanical mode switch settings 1 and 4 are required.
Mechanical entanglement is generated by choosing the switch settings 2 and 4 and homodyning the optical output, which corresponds to Bayesian inference of a joint mechanical quadrature.
In the verification procedure, the same combination of paths as in the precooling stage will yield the blocks of the covariance matrix corresponding to each mechanical oscillator, while the switch  setting 2 and 4 allows the mechanical-mechanical correlations to be accessed.
}
\label{nonintsetup}
\end{figure}

%Explanation and decoherence map
The non-interferometric scheme for entanglement preparation and verification using pulsed optomechanics is shown in the circuit diagram of Fig.~\ref{nonintsetup}. As before, a precooling stage is implemented to increase the amount of mechanical entanglement generated. The entangling stage is then initiated by a pulse of coherent light entering an optomechanical cavity and interacting with the first mechanical mode---identified by its quadrature vector $\mathbf{X}_{\textsc{m}\scriptscriptstyle{1}}$. The light is then subject to optical losses, each described by $\mathcal{E}_{\eta_{\textsc{cav}}}$, corresponding to the output-cavity-coupling and input-cavity-coupling efficiencies of the first and second cavities, respectively. Here, we assume equal outcoupling and incoupling efficiencies, which are given by $\eta_{\textsc{cav}}$. An optional optical squeezer may be placed between these two optical loss channels to provide resilience to interactions with the optical environment.

For the first optomechanical interaction, it was sufficient to absorb the input-cavity-coupling losses into the definition of the interaction strength $\chi$, but when the light interacts with the second mechanical mode---described by $\mathbf{X}_{\textsc{m}\scriptscriptstyle{2}}$---it is important to separate out this effect. This is because the light incident on the second cavity contains information about $\XMo$, and so input coupling losses lead to loss of information about the mechanical position and a reduction in the final amount of entanglement generated. The attenuation of the light between the two optomechanical interactions leads to a reduction in the optomechanical interaction strength: $\chi\rightarrow\eta_{\textsc{cav}}\chi$. Hence the second optomechanical interaction is modified from $U_{\textsc{om}}=\rme^{\rmi\chi\XL\XM}$ to $U_{\textsc{om}}'=\rme^{\rmi\eta_{\textsc{cav}}\chi\XL\XM}$.

After the optomechanical interaction with the second mechanical oscillator, the light passes through the optical loss channels $\mathcal{E}_{\eta_{\textsc{cav}}}$ and $\mathcal{E}_{\eta_{\textsc{det}}}$, with an optional squeezing operation in between. Finally, a homodyne measurement is made on the phase quadrature of light which contains information about an EPR-like quadrature variable of $\XMo$ and $\XMt$. Hence, this measurement projects the mechanical subspace into an entangled state via Bayesian inference of a joint-mechanical quadrature---see \ref{MeasOpApproach} for a more complete discussion. This non-interferometric entanglement protocol, followed by free mechanical evolution $U_{\textsc{rot}}(\theta,\phi)$ and mechanical decoherence $\mathcal{E}_{\gamma}$, is represented by Eq.~\eqref{nonintmap}.
\begin{gather}\label{nonintmap}
   \rho\rightarrow \mathcal{E}_{\gamma}(U_{\textsc{rot}}(\theta,\phi)\ket{\PL}\bra{\PL}\circ\mathcal{E}_{\eta_{\textsc{det}}}(U_{\textsc{sq}}\circ\\\nonumber
   \mathcal{E}_{\eta_{\textsc{cav}}}(U_{\textsc{om}}'\circ\mathcal{E}_{\eta_{\textsc{cav}}}(U_{\textsc{sq}}\circ\mathcal{E}_{\eta_{\textsc{cav}}}(U_{\textsc{om}}\circ\rho))))).
\end{gather}

%setup

%verification

%plots
\begin{figure}
\vspace{-3cm}
\includegraphics[width=0.89167\linewidth,angle=270]{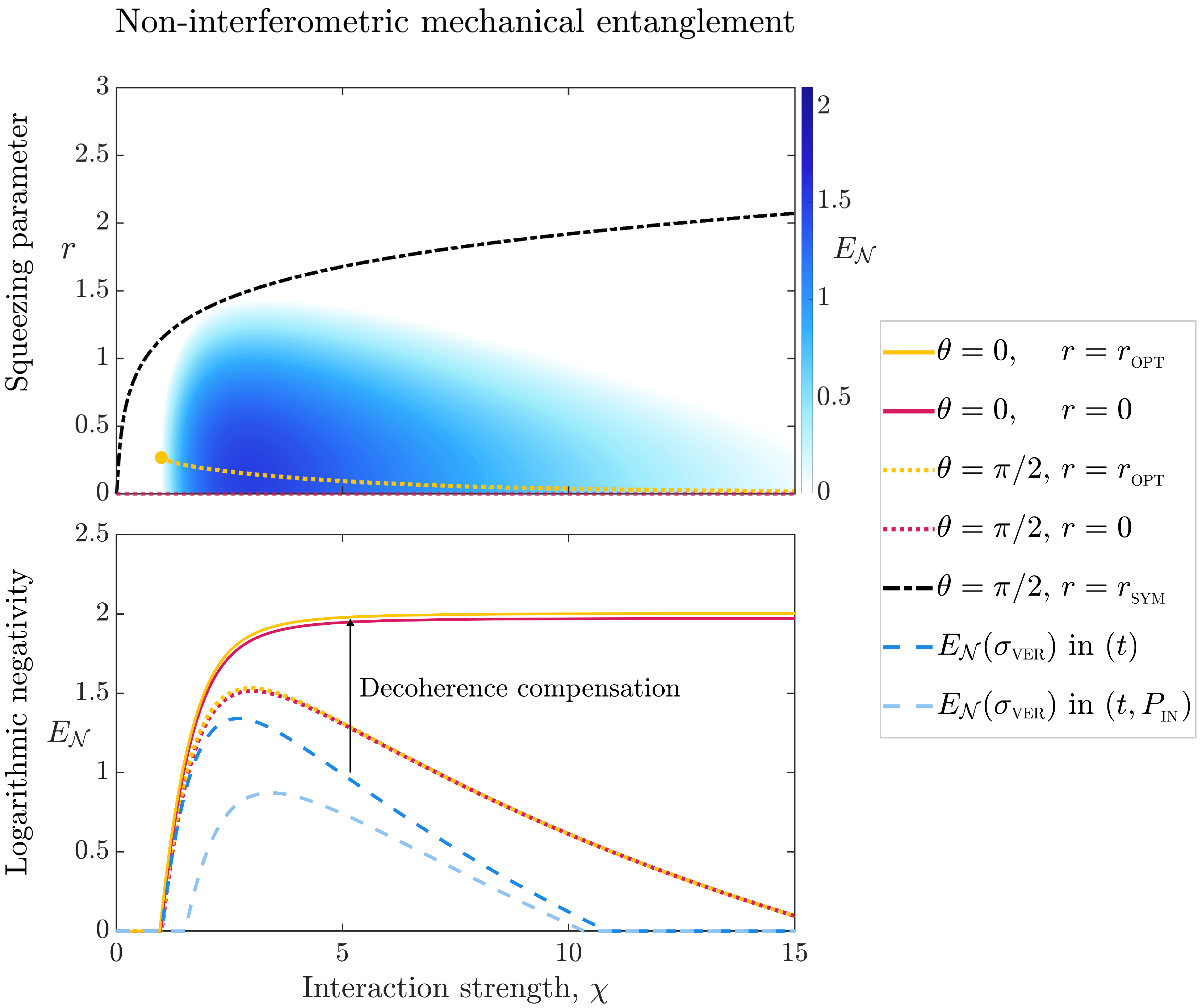}
\caption{ \textbf{Logarithmic negativity $E_{\mathcal{N}}(\sigma)$ of the entangled mechanical state generated by our non-interferometric scheme.} \textbf{Upper:} Contour plot of logarithmic negativity $E_{\mathcal{N}}(\sigma)$ as a function of optomechanical interaction strength $\chi$ and squeezing parameter $r$---with other system parameters corresponding to those listed in Table \ref{parameters}. Here, each mechanical mode has decohered over a quarter of a period since the time of entanglement generation. The maximum of the logarithmic negativity is $E_{\mathcal{N}}=1.53$, which corresponds to $\chi= 2.97$ and $r=0.15$. Without the use of an optical squeezer the protocol achieves  $E_{\mathcal{N}}=1.51$ with $\chi=3.00$.  As compared to the previous entanglement schemes, the asymmetry in the optomechanical interactions leads to a different form for $r_{\textsc{sym}}$, which is given by the real-positive solution of a quartic equation in $\rme^{2r}$. Also in contrast to the previous schemes, the first point in the contour plot with $E_{\mathcal{N}}(\sigma)>0$
corresponds to finite value of $r_{\textsc{opt}}$, which we highlight with the yellow circle. \textbf{Lower:} Logarithmic negativity of the entangled state, showing the effect of mechanical decoherence and optical squeezing on $E_{\mathcal{N}}(\sigma)$, and conservative approaches for mechanical state verification. As the mechanical modes evolve and undergo decoherence, the optical squeezing operations become less useful---demonstrated by the convergence of the orange and red curves. In addition to conservative approaches, the physical requirement that $\phi\geq\theta$ leads to further decay of the elements of $\sigma_{\textsc{ver}}$.}
\label{nonintlogneg}
\end{figure}

Focusing on the verification procedure, the $A_{\textsc{ver}}$ and $B_{\textsc{ver}}$ blocks of the verified covariance matrix $\sigma_{\textsc{ver}}$ are obtained by addressing each mechanical oscillator individually with a pulse of light, which is achieved by the use of optical switches as shown in Fig.~\ref{nonintsetup}. The $A_{\textsc{ver}}$ and $B_{\textsc{ver}}$ blocks are therefore given by an equivalent expression to Eq.~\eqref{verB}. A single verification pulse, with an initial phase quadrature $P_{\textsc{in}}$, is then used to construct $C_{\textsc{ver}}$. By allowing both mechanical modes to evolve over a phase-space angle $\theta$ and implementing a controllable delay line between the two mechanical modes, such that the second mechanical oscillator evolves over a total phase-space angle $\phi$, the phase quadrature entering the homodyne detector is given by
$\Pv(\theta,\phi)=P_{\textsc{in}}+\chi(\XMo(\theta)+\XMt(\phi))$. Due to the non-interferometric arrangement, we have the condition $\phi\geq\theta$, which leads to further decay of the $C_{\textsc{ver}}$-elements---on top of the decay due to \textit{conservative in time} approach. By defining $\epsilon(\theta,\phi)=	\mathrm{Var}(P_{\textsc{v}}(\theta, \phi)) - \mathrm{Var}(P_{\textsc{v}}(\theta, \phi+\pi))$, we find that the elements describing the mechanical correlations are given by
\begin{equation} \label{CBlocknonint}
C_{\textsc{ver}} = \frac{1}{4\chi^2} \begin{pmatrix}
	\epsilon(2\pi,2\pi) & \epsilon(2\pi,\frac{5\pi}{2})  \\
	-\epsilon(\frac{5\pi}{2},3\pi) & \epsilon(\frac{5\pi}{2},\frac{5\pi}{2})\\
		\end{pmatrix}.
\end{equation}

Plots analysing the behaviour of the logarithmic negativity obtained using the non-interferometric scheme are shown in Fig.~\ref{nonintlogneg}, and we see that the protocol is relatively insensitive to the squeezing operations as compared to the optical-mechanical and interferometric entanglement schemes. 
This difference in sensitivity occurs because there is an additional optomechanical interaction and loss channel which further modifies the same optical phase-space distribution towards the ideal, slightly asymmetric, state with $\mathrm{Var}(\PL)>\mathrm{Var}(\XL)$. This additional modification to the phase-space distribution means a lower squeezing parameter $r$ is needed to protect the state from optical loss and mechanical decoherence. The lower plot of Fig.~\ref{nonintlogneg} shows the logarithmic negativity which may be measured in the verification process, compared to the logarithmic negativity of the unverified state at different times since entanglement generation.

\section{Comparison between the entanglement schemes}\label{comparisonsection}
In this section, we compare the three entanglement schemes outlined above. In particular, we discuss the relative importance of the squeezing operations, the fidelity of the verification procedure, and the sensitivity to the cavity coupling efficiency. Focus will be placed primarily on the interferometric and non-interferometric mechanical entanglement schemes to establish which one is favourable for preparing and verifying entangled mechanical states.

\begin{figure}
\vspace{-2.5cm}\hspace{-1cm}\includegraphics[width=0.45727\linewidth,angle=270]{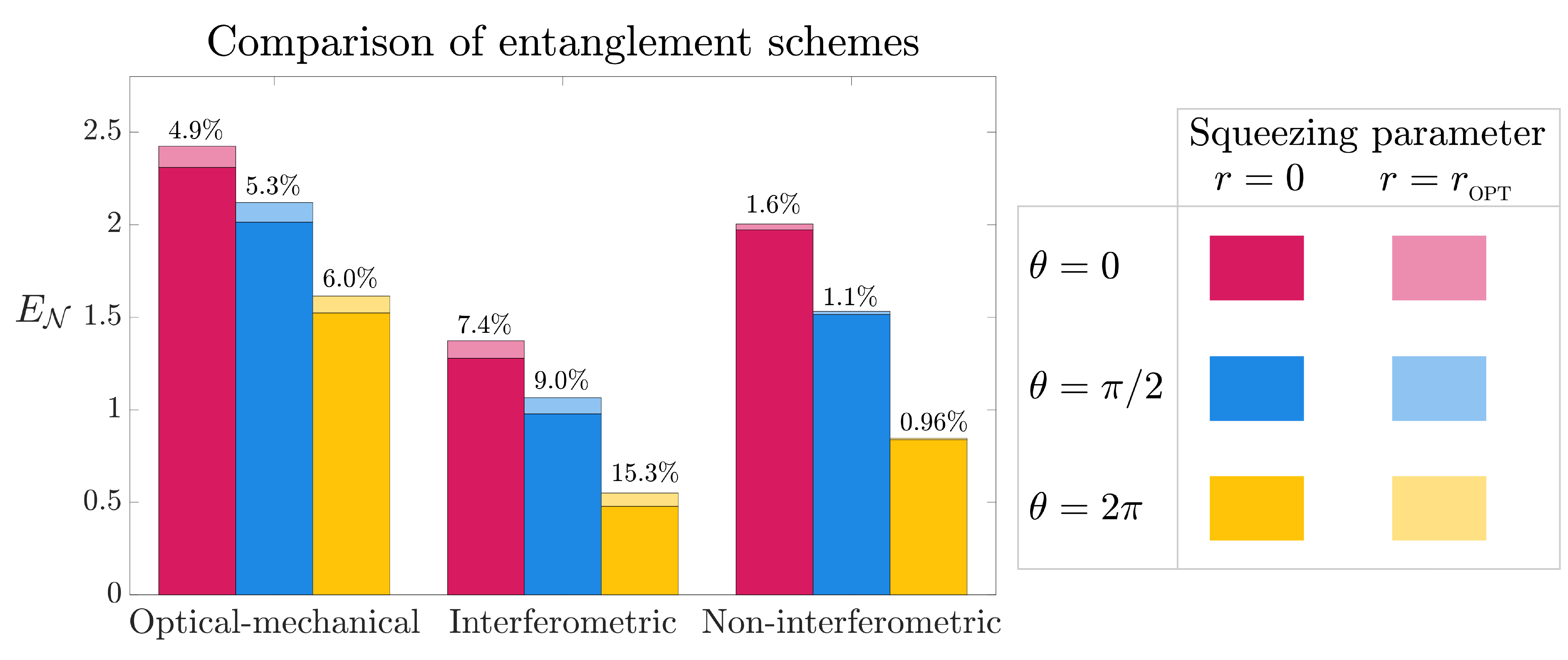}
\caption{\textbf{Comparison of entanglement schemes with and without optical squeezers.} The maximum logarithmic negativity $E_{\mathcal{N}}(\sigma)$ generated using the three different entanglement schemes: optical-mechanical, interferometric, and non-interferometric. The squeezing parameter is taken to either be $r=0$ or the value which maximizes logarithmic negativity $r=r_{\textsc{opt}}$. The phase-space angle $\theta$ over which the mechanical modes rotate and decohere, after the time of entanglement generation, is taken to be $0$, $\pi/2$, or $2\pi$. For the mechanical entanglement schemes, we take the second mechanical phase-space angle to be $\phi=\theta$. The percentage increase in $E_{\mathcal{N}}(\sigma)$, as a result of optimal squeezing, is listed above each bar.}
\label{barplots}
\end{figure}

%Squeezing
In Fig.~\ref{barplots}, we compare the logarithmic negativity generated in each of the three entanglement schemes and also analyze the relative importance of optical squeezing in each scheme. More specifically, in this figure, we plot the percentage increase in the logarithmic negativity due to the use of optical squeezers, and find that squeezing is most useful in the interferometric scheme and least useful in the non-interferometric scheme. It is interesting to note that as the mechanical oscillators evolve and decohere after the instance of entanglement generation, the percentage increase in logarithmic negativity rises for the optical-mechanical and interferometric entanglement schemes, while the percentage increase drops towards $0\%$ for the non-interferometric scheme. This is due to the active role the second optomechanical interaction has on the optical phase-space distribution in the non-interferometric scheme. For the parameter values listed in Table \ref{parameters}, we establish that a squeezing parameter of $r\approx0.60$ is required to reach maximum logarithmic negativity for the optical-mechanical and interferometric entanglement protocols, which is experimentally feasible. However, it is, of course, experimentally simpler to implement the entanglement protocol without the use of optical squeezers, in which case the non-interferometric scheme is advantageous.

In the conservative approaches for state verification, the $C_{\textsc{ver}}$ elements in the non-inteferometric protocol must be measured at later times than those in the interferometric protocol. Despite this requirement, the logarithmic negativity of the verified state $E_{\mathcal{N}}(\sigma_{\textsc{ver}})$ is much higher in the non-interferometric scheme than in the interferometric scheme---as is evident from the lower plots of Figs.~\ref{intlogneg} and \ref{nonintlogneg}. Specifically, in the interferometric scheme, the conservative in time approach reaches a maximum logarithmic negativity of $E_{\mathcal{N}}(\sigma_{\textsc{ver}})=0.90$, as compared to the non-interferometric scheme which reaches $E_{\mathcal{N}}(\sigma_{\textsc{ver}})=1.34$. 

To aid experimental development, here we now investigate the behaviour of the logarithmic negativity as the cavity coupling efficiency is varied, keeping all other parameters in Table \ref{parameters} the same. Furthermore, we also set all the squeezing parameters to $r=0$ for this analysis and assume that the input-cavity-coupling and output-cavity-coupling efficiencies are both given by $\eta_{\textsc{cav}}$. We are therefore interested in finding the minimum value of $\eta_{\textsc{cav}}$ required to satisfy the condition $E_{\mathcal{N}}(\sigma)>0$ in each of the three schemes. Table \ref{minetatable} lists this minimum value of $\eta_{\textsc{cav}}$ for different values of the phase-space angle through which the mechanical modes evolve and decohere after entanglement generation. We find that the optical-mechanical system can still be entangled, even after the mechanical mode has decohered over a full period, for $\eta_{\textsc{cav}}>0.0065$. This entanglement scheme is therefore currently experimentally feasible. The mechanical entanglement schemes lead to stricter requirements on the cavity coupling efficiency, with the non-interferometric scheme being favourable in this regard. However, if one calculates the total optical efficiency in each of the two schemes, we find that only relatively low efficiencies are required to generate entanglement. More specifically, by using Table \ref{minetatable} and Eqs.~\eqref{DMMap} and \eqref{nonintmap}, one can calculate that in order to generate entanglement total optical efficiencies of $50\%$ and $11\%$ are required for the interferometric and non-interferometric schemes, respectively. Here, the total optical efficiency in the interferometric design is given by the efficiency of a single optical path, i.e. the product of the cavity coupling efficiency and the detection efficiency, while for the linear non-interferometric design the total efficiency is given by product of all the intensity efficiencies along the optical path. Moreover, Table \ref{minetatable} also shows the minimum cavity coupling efficiency required to verify entanglement in each of the schemes, i.e. to satisfy $E_{\mathcal{N}}(\sigma_{\textsc{ver}})>0$ in the case of no squeezing. These values correspond to minimum total optical efficiencies of $0.63\%$, $60\%$, and $16\%$ to verify entanglement in the optical-mechanical, interferometric, and non-interferometric schemes, respectively.

\begin{table}[h]
 \caption{\textbf{Minimum cavity coupling efficiency for entanglement generation and verification.} The table shows the minimum cavity coupling efficiency ${\eta_{\textsc{cav}}}$ required to retain an entangled state after each mechanical oscillator rotates through a phase-space angle $\theta$ whilst undergoing some decoherence, and the minimum cavity coupling efficiency required to verify the entangled state using the conservative in time approach. Other systems parameters correspond to those given in Table \ref{parameters}, apart from the optical squeezing parameter $r$---which we set to zero here. The total optical efficiency is given by the product of the intensity efficiencies along the optical path.}
\centering
\begin{tabular}{c|c c c c}
\hline\hline
Entanglement scheme & $\theta=0$ & $\theta=\pi/2$ & $\theta=2\pi$ &Verification \\
\hline
    Optical-mechanical & $>$0 & 0.00040 & 0.0065 & 0.0066 \\
    Interferometric & 0.53 & 0.59 & 0.70 & 0.63 \\
    Non-interferometric & 0.48 & 0.52 & 0.61 &0.55 \\
    
     \hline\hline
     \end{tabular}
     \label{minetatable}
     \end{table}

To summarise this comparison between schemes, in the parameter range we consider, the non-interferometric design allows for both the preparation and verification of a higher amount of mechanical entanglement---both with and without the use of optical squeezers. The non-interferometric scheme also provides a route for generating entangled mechanical states if the cavity coupling efficiency is the major source of technical difficulty. In fact, the only regime we have studied in which the interferometric design is preferable, is when the output coupling efficiency from the cavities is set to unity, but the input coupling efficiency remains less than unity. In this case, the non-monotonic behaviour in the logarithmic negativity with increasing interaction strength $\chi$ is removed as the optical squeezers act before any decoherence channels, and so the entangled state may then be made resilient to optical loss and mechanical decoherence---see \ref{nonmonappendix} for a detailed discussion. However, if both input and output coupling efficiencies are set to unity, then the non-interferometric scheme allows for higher values of logarithmic negativity to be reached, and also allows the non-monotonic behaviour to be avoided.

\section{Conclusion}
In this work, we have introduced schemes for the preparation and verification of optical-mechanical and mechanical-mechanical entanglement in the unresolved-sideband regime of optomechanics. These schemes utilize short pulsed interactions and optical homodyne measurements to generate these types of entanglement, and by using optical squeezers and an additional precooling stage, we see an increase in the entanglement generated in the presence of both mechanical decoherence and optical loss. Moreover, we have also carefully studied the problem of state verification and have designed procedures for accessing each element of the covariance matrices, while respecting the requirement that each element must be measured at a specific time. 

Two mechanical entanglement schemes are introduced and compared using a feasible set of parameters based on current state-of-the-art experiments. We concluded that the non-interferometric design is favourable to the interferometric design in several ways. Whilst both schemes are capable of generating a large amount of entanglement---i.e. $E_{\mathcal{N}}>1$, the non-interferometric scheme is able to prepare and verify more entanglement in the parameter regime we consider. Furthermore, mechanical entanglement can be prepared and verified at a minimum total optical efficiency of $16\%$ using the non-interferometric approach even without squeezing, while the interferometric scheme requires a higher optical efficiency and squeezing parameter. However, we see that preparing and verifying optical-mechanical entanglement can even be achieved with optical efficiencies much less than $1\%$ and so can be realized with present-day experiments.

This work provides a promising path for the development of new quantum technologies, entangled resources for quantum-information-based tasks, and for the exploration of the foundations of physics using the tools of quantum optics.

\section*{Acknowledgements}
We would like to thank John Price and Rufus Clarke for valuable discussions. This project was supported by the Engineering and Physical Sciences Research Council (EP/N014995/1), UK Research and Innovation (MR/S032924/1), The Branco Weiss Fellowship---Society in Science (administered by the ETH Zurich), and the Royal Society (via a Research Grant and a Wolfson Research Merit Award). MSK also acknowlegdges funding from the QuantERA ERA-NET Cofund in Quantum Technologies implemented within the European Union's Horizon 2020 Programme.

\newpage
\appendix
\section{Symplectic transformations, Gaussian operations, and entangled state preparation in phase space}\label{symplectictransformations}
\subsection{Symplectic transformations}

In the Heisenberg picture, the action of a Gaussian unitary $U_{\textsc{G}}$ on the $2n$-dimensional quadrature vector $\mathbf{X}$ is described by the equivalent transformation $S\mathbf{X}$.  Here, $S$ belongs to the real symplectic group---$S\in{Sp(2n,\mathds{R})}$. Therefore, under the action of a Gaussian unitary operation, the first moments and the covariance matrix of a state transform as follows:
\begin{align}
    \braket{\Xvec} \rightarrow S \braket{\Xvec} \\
    \sigma \rightarrow S \sigma S^\mathrm{T}. 
\end{align}
A necessary and sufficient condition for the covariance matrix $\sigma$ to represent a Gaussian state is given by the uncertainty relation $\sigma+\rmi\Omega/2\geq0$~\cite{Simon1994}. As the symplectic matrix $S$ preserves the symplectic form $S\Omega{S}^{\mathrm{T}}=\Omega$, the symplectic transform preserves the Gaussian character of the quantum state.

\subsubsection{Single-mode transformation}
Consider a single mode described by the quadrature vector $\mathbf{X}=(X,P)^{\mathrm{T}}$, with $X=(a+a^{\dagger})/\sqrt{2}$ and $P=-\rmi(a-a^{\dagger})/\sqrt{2}$. Below, we list the symplectic single-mode transformations used in this work.

\textit{Rotation}--- Evolution over time $t=\theta/\omega$ under the free Hamiltonian $H=\hbar\omega{a}^{\dagger}a$, corresponds to a rotation by an angle $\theta$ in phase space, and has the symplectic matrix
\begin{equation}
S_{\textsc{rot}}(\theta)=\begin{pmatrix}\cos\theta &\sin\theta\\
-\sin\theta&\cos\theta\end{pmatrix}.
\end{equation}

\textit{Squeezer}--- A squeezing operation is described by the unitary operator
$U_{\textsc{sq}}=\exp{\frac{1}{2}[r({a}^{\dagger})^2-r^{*}{a}^2]}$. For $r\in\mathds{R}$, the corresponding symplectic matrix is given by
\begin{equation}
S_{\textsc{sq}}(r)=\begin{pmatrix}\rme^{r} &0\\
0&\rme^{-r}\end{pmatrix}.
\end{equation}

\subsubsection{Two-mode transformations}
A two-mode quadrature vector may be written as $\mathbf{X}=(X_{1},P_{1},X_{2},P_{2})^{\mathrm{T}}$. Here, $X_{i}=(a_{i}+a_{i}^{\dagger})/\sqrt{2}$, $P_{i}=-\rmi(a_{i}-a_{i}^{\dagger})/\sqrt{2}$, and $i=1,2$.

\textit{Beamsplitter}--- The unitary operation $U_{\textsc{bs}}=\mathrm{exp}\{\alpha\cos\beta(a^{\dagger}_{1}a_{2}-a_{1}a_{2}^{\dagger})+\rmi\alpha\sin\beta(a^{\dagger}_{1}a_{2}+a_{1}a_{2}^{\dagger})\}$ acts as a general beamsplitter operation and corresponds to the symplectic transformation
\begin{equation}
S_{\textsc{bs}}(\alpha,\beta)=\begin{pmatrix}
\cos\alpha & 0 & \sin\alpha\cos\beta & -\sin\alpha\sin\beta \\
0 & \cos\alpha & \sin\alpha\sin\beta & \sin\alpha\cos\beta \\
-\sin\alpha\cos\beta & -\sin\alpha\sin\beta & \cos\alpha & 0 \\
\sin\alpha\sin\beta & -\sin\alpha\cos\beta & 0 & \cos\alpha
\end{pmatrix}.
\end{equation}

\textit{Linearized-pulsed optomechanical interaction}--- The optomechanical interaction described by $\rme^{\rmi\chi{X}_{1}X_{2}}$ corresponds to the symplectic transformation
\begin{eqnarray}
S_{\textsc{om}}=\begin{pmatrix}
1&0&0&0\\
0&1&\chi&0\\
0&0&1&0\\
\chi&0&0&1
\end{pmatrix}.
\end{eqnarray}

\subsubsection{Decoherence}
A phase-insensitive Gaussian decoherence channel acts on the quantum state $\rho$ according to $\mathcal{E}_{\textsc{g}}(\rho)$. The channel $\mathcal{E}_{\textsc{g}}(\rho)$ is the result of carrying out a trace operation on a phase-insensitive bath after a system-bath unitary interaction. At the level of the first moments vector $\braket{\Xvec}$ and covariance matrix $\sigma$, the corresponding phase-space mapping is of the form~\cite{Paris2005}
\begin{align}
   \braket{\Xvec} &\rightarrow G^{1/2}\braket{\Xvec}\\
    \sigma\rightarrow G^{1/2} \sigma &G^{1/2} + (\mathds{1} - G) \sigma_{\mathrm{env}}\label{covdecomap}.
\end{align}
This map allows one to model both optical and mechanical decoherence in the entanglement protocols. 
The description of the optical loss channel $\mathcal{E}_{\eta}$ may be achieved by setting $G = \eta \mathds{1}_{2}$ and $\sigma_{\mathrm{env}}$ to the vacuum optical state. Here, $\eta$ is the intensity efficiency of the loss channel. 
Describing the mechanical decoherence channel $\mathcal{E}_{\gamma}$ is accomplished by letting  $G = \rme^{- \gamma t}  \mathds{1}_{2}$ and $\sigma_{\mathrm{env}}$ be a thermal state with mean occupation $\bar{N}$. Here, the decoherence channel acts on the mechanical modes for a time $t$ and $\gamma$ is the intrinsic mechanical decay rate.

\subsubsection{Measurements}

In Hilbert space, we may describe a Gaussian measurement with the Krauss operator $K_{\textsc{g}}$, such that after the measurement the state $\rho$ transforms to a state proportional to $K_{\textsc{g}}\circ\rho$. By considering the first-moments vector $\braket{\textbf{X}}$ and the covariance matrix of Eq.~\eqref{blockCM}, a Gaussian measurement on the optical mode can be described through the mapping
\begin{align}
\label{MeasEquationappendix}
    \braket{\mathbf{\XM}} &\rightarrow\braket{\mathbf{\XM}}+C^{\mathrm{T}}(A+\sigma_{\mathrm{meas}})^{-1}(\braket{\Xvec_{\mathrm{meas}}}-\braket{\mathbf{\XL}}),\\
    B &\rightarrow{B}-C^{\mathrm{T}}(A+\sigma_{\mathrm{meas}})^{-1}C\label{covMeasEquationappendix}.
\end{align}
Here, $\braket{\textbf{X}_{\mathrm{meas}}}$ corresponds to the measurement outcome and $\sigma_{\mathrm{meas}}$ describes the Gaussian measurement~\cite{eisert2002distilling}. In the limit of a projection onto an infinitely squeezed Gaussian state, which constitutes an optical homodyne measurement, the term $(A+\sigma_{\mathrm{meas}})^{-1}$ becomes $(\Pi_{\varphi}A\Pi_{\varphi})^{\textsc{MP}}$. Here, $\mathrm{MP}$ refers to the Moore-Penrose pseudo-inverse and $\Pi_{\varphi}$ is the projection operator onto the $\XL(\varphi)$ quadrature
\begin{equation}
 \Pi_{{\varphi}}=\left(\begin{array}{c c}
\cos^2\varphi & \cos\varphi\sin\varphi \\
\cos\varphi\sin\varphi & \sin^2\varphi
\end{array}\right).
\end{equation}
The Krauss operator corresponding to $\Pi_{\varphi}$ is given by the Hilbert-space projector $\ket{\XL(\varphi)}\bra{\XL(\varphi)}$, with $\int_{-\infty}^{+\infty}\rmd\XL(\varphi)\ket{\XL(\varphi)}\bra{\XL(\varphi)}=\mathds{1}$.

\subsection{Precooling stage}
Eq.~\eqref{omintmeas} describes the precooling stage of the mechanical oscillators. The symplectic formalism outlined above may be used to neatly compute the analytic result of this map. Namely, if the initial state of the optical-mechanical system is given by the covariance matrix
\begin{equation}
    \sigma_{\textsc{in}}= \frac{1}{2}\begin{pmatrix}
   1&0&0&0\\
   0&1&0&0\\
   0&0&1+2\bar{n}&0\\
   0&0&0&1+2\bar{n}
    \end{pmatrix},
\end{equation}
where $\bar{n}$ is the initial mechanical occupation, then the precooled state of the mechanics is given by
\begin{equation}
    \sigma_{\textsc{cool}}= \begin{pmatrix}
   V_{\textsc{x}}&0\\
   0&V_{\textsc{p}}
    \end{pmatrix}.
\end{equation}
Here,
\begin{eqnarray}
2V_{\textsc{x}}&=&1+2\bar{N}+2\rme^{-\gamma t}(\bar{n}-\bar{N})+\rme^{-\gamma t}\chi^2 \\
2V_{\textsc{p}}&=&(1+2\bar{N})(1-\rme^{-\gamma t})+\rme^{-\gamma t}\dfrac{1+2\bar{n}}{1+\eta\chi^2(1+2\bar{n})}.
\end{eqnarray}

\subsection{Optical-mechanical entanglement scheme}
The preparation of an entangled state of light and a precooled mechanical oscillator, followed by optical and mechanical decoherence, is described by Eq.~\eqref{omintsq}. Taking $\frac{1}{2}\mathds{1}_{2}\oplus\sigma_{\textsc{cool}}$ as the initial covariance matrix of light and mechanics, then the entangled state has the covariance matrix 
\begin{eqnarray}
 \sigma_{\textsc{lm}}&=&\begin{pmatrix}
 A_{\textsc{lm}}&&C_{\textsc{lm}}\\
  C_{\textsc{lm}}^{\mathrm{T}}&&B_{\textsc{lm}}
\end{pmatrix},
\end{eqnarray}
with
\begin{eqnarray}
 A_{\textsc{lm}}&=&\begin{pmatrix}
 a_{11}&0\\0&a_{22}
 \end{pmatrix},\\
  B_{\textsc{lm}}&=&\begin{pmatrix}
 b_{11}&0\\0&b_{22}
 \end{pmatrix},\\
 C_{\textsc{lm}}&=&\begin{pmatrix}
0&c_{12}\\ c_{21}&0
 \end{pmatrix},
\end{eqnarray}
where
\begin{eqnarray}
2a_{11}&=&1-\eta_{\textsc{det}}+\rme^{2r}\eta_{\textsc{det}}\\
2a_{22}&=&1-\eta_{\textsc{det}}+\rme^{-2r}\eta_{\textsc{det}}(1+2V_{\textsc{x}}\eta_{\textsc{cav}}\chi^2)\\
2b_{11}&=&(1+2\bar{N})(1-\rme^{-\gamma t})+2V_{\textsc{x}}\rme^{-\gamma t}\\
2b_{22}&=&(1+2\bar{N})(1-\rme^{-\gamma t})+(2V_{\textsc{p}}+\chi^2)\rme^{-\gamma t}\\
2c_{12}&=&\rme^{r}\rme^{-\gamma t/2}\sqrt{\eta}\chi\\
2c_{21}&=&2V_{\textsc{x}}\rme^{-r}\rme^{-\gamma t/2}\sqrt{\eta}\chi.
\end{eqnarray}
For brevity, $\sigma_{\textsc{lm}}$ is presented in a frame rotating at the mechanical frequency, and we will also present the mechanical entanglement protocols in the same rotating frame. However, as the decoherence channel $\mathcal{E}_{\gamma}$ and the unitary operation described by $U_{\textsc{rot}}$ commute, the form of $\sigma_{\textsc{lm}}$ in a non-rotating frame may be obtained by computing $\mathds{1}_{2}\oplus{S}_{\textsc{rot}}(\theta) \sigma_{\textsc{lm}}\mathds{1}_{2}\oplus{S}_{\textsc{rot}}^{\mathrm{T}}(\theta)$.

\subsection{Interferometric mechanical entanglement scheme}

The protocol that generates an entangled state of two mechanical oscillators using the interferometric scheme is described by Eq.~\eqref{DMMap}. If the initial state of each mechanical oscillator is described by the covariance matrix $\sigma_{\textsc{cool}}$, the covariance matrix of each light mode interacting with the mechanical modes is $\frac{1}{2}\mathds{1}_{2}$, and the homodyne angles are chosen to be $(\varphi,\psi)=(0,\pi/2)$, the final bipartite mechanical covariance matrix is given by
\begin{eqnarray}
\sigma_{\textsc{int}}&=&\begin{pmatrix}
 A_{\textsc{int}}&&C_{\textsc{int}}\\
  C_{\textsc{int}}^{\mathrm{T}}&&B_{\textsc{int}}
\end{pmatrix},\\
 A_{\textsc{int}}&=&B_{\textsc{int}}=\begin{pmatrix}
 a_{11}&0\\0&a_{22}
 \end{pmatrix},\\
 C_{\textsc{int}}&=&\begin{pmatrix}
 c_{11}&0\\0&c_{22}
 \end{pmatrix},
\end{eqnarray}
where
\begin{eqnarray}
2a_{11}&=&(1+2\bar{N})(1-\rme^{-\gamma t})+2V_{\textsc{x}}\rme^{-\gamma t}\left(1-
\frac{\eta{V}_{\textsc{x}}\chi^2}{\rme^{2r}(1-\eta_{\textsc{det}})+\eta_{\textsc{det}}(1+2V_{\textsc{x}}\eta_{\textsc{cav}}\chi^2)}\right)\nonumber,\\
2a_{22}&=&(1+2\bar{N})(1-\rme^{-\gamma t})+\rme^{-\gamma t}\left[2V_{\textsc{p}}+\chi^2-\frac{1}{2}\left(\frac{\rme^{2r}\eta\chi^2}{1+\eta_{\textsc{det}}(\rme^{2r}-1)}\right)\right],\nonumber\\
c_{11}&=&\frac{\rme^{-\gamma t}\eta{V}_{\textsc{x}}^2\chi^2}{\rme^{2r}(1-\eta_{\textsc{det}})+\eta_{\textsc{det}}(1+2V_{\textsc{x}}\eta_{\textsc{cav}}\chi^2)},\nonumber\\
c_{22}&=&-\frac{1}{4}\left(\frac{\rme^{2r}\rme^{-\gamma t}\eta\chi^2}{1+\eta_{\textsc{det}}(\rme^{2r}-1)}\right).\nonumber
\end{eqnarray}
Note that as $A_{\textsc{int}}=B_{\textsc{int}}$, the state is symmetric and so, in addition to the logarithmic negativity, the entanglement of formation may be also computed.

\subsection{Non-interferometric mechanical entanglement scheme}
Eq.~\eqref{nonintmap} describes the generation of an entangled state of two mechanical oscillators using the non-interferometric scheme. As in the previous subsection, if the initial state of each mechanical oscillator is described by the covariance matrix $\sigma_{\textsc{cool}}$, and the initial covariance matrix of the light mode is $\frac{1}{2}\mathds{1}_{2}$, then the final bipartite mechanical covariance matrix is 
\begin{eqnarray}
\sigma_{\textsc{non}}&=&\begin{pmatrix}
 A_{\textsc{non}}&&C_{\textsc{non}}\\
  C_{\textsc{non}}^{\mathrm{T}}&&B_{\textsc{non}}
\end{pmatrix},\\
 A_{\textsc{non}}&=&\begin{pmatrix}
 a_{11}&0\\0&a_{22}
 \end{pmatrix},\\
  B_{\textsc{non}}&=&\begin{pmatrix}
 b_{11}&0\\0&b_{22}
 \end{pmatrix},\\
 C_{\textsc{non}}&=&\begin{pmatrix}
 c_{11}&0\\0&c_{22}
 \end{pmatrix},
\end{eqnarray}
where
\begin{eqnarray*}
2a_{11}&=&(1+2\bar{N})(1-\rme^{-\gamma t})+\rme^{-\gamma t}V_{\textsc{x}}J_{a}\\
2a_{22}&=&(1+2\bar{N})(1-\rme^{-\gamma t})+\rme^{-\gamma t}(2V_{\textsc{p}}+\chi^2)\\
2b_{11}&=&(1+2\bar{N})(1-\rme^{-\gamma t})+\rme^{-\gamma t}V_{\textsc{x}}J_{b}\\
2b_{22}&=&(1+2\bar{N})(1-\rme^{-\gamma t})+\rme^{-\gamma t}\left[2V_{\textsc{p}}+\eta_{\textsc{cav}}^2\chi^2[1+\eta_{\textsc{cav}}(\rme^{2r}-1)]\right]\\
c_{11}&=&-\frac{1}{8}\rme^{2r}V_{\textsc{x}}^2\rme^{-\gamma t}\eta_{\textsc{cav}}^3\eta_{\textsc{det}}\chi^2J_{c}\\
c_{22}&=&\frac{1}{2}\rme^{r}\rme^{-\gamma t}\eta_{\textsc{cav}}^2\chi^2.
\end{eqnarray*}
The functions $J_{a}$, $J_{b}$, and $J_{c}$ are given by
\begin{eqnarray*}
    J_{a}&=&{\bigg\{}\rme^{8r}(1-\eta_{\textsc{det}})^2+2\rme^{6r}\eta_{\textsc{det}}(1-\eta_{\textsc{det}})(1-\eta_{\textsc{cav}}^2+2V_{\textsc{x}}\eta_{\textsc{cav}}^3\chi^2)\\
    &+&\rme^{4r}\eta_{\textsc{det}}\left\{\eta_{\textsc{det}}+\eta_{\textsc{cav}}^2\left[2-(4-\eta_{\textsc{cav}}^2)\eta_{\textsc{det}}+2V_{\textsc{x}}\eta_{\textsc{cav}}(1+(1-2\eta_{\textsc{cav}}^2)\eta_{\textsc{det}})\chi^2
    +4V_{\textsc{x}}^2\eta_{\textsc{cav}}^4\eta_{\textsc{det}}\chi^4
    \right]
    \right\}\\
    &+&\rme^{2r}\eta_{\textsc{cav}}^2\eta_{\textsc{det}}^2\left\{2+\eta_{\textsc{cav}}[2V_{\textsc{x}}\chi^2-2\eta_{\textsc{cav}}(1-V_{\textsc{x}}\eta_{\textsc{cav}}\chi^2(1+2V_{\textsc{x}}\eta_{\textsc{cav}}\chi^2))
    ]
    \right\}\\
    &+&\eta_{\textsc{cav}}^4\eta_{\textsc{det}}^2[1+2V_{\textsc{x}}\eta_{\textsc{cav}}\chi^2] {\bigg\}}\bigg/{\bigg\{}\rme^{8r}(1-\eta_{\textsc{det}})^2\\
    &+&2\rme^{6r}\eta_{\textsc{det}}(1-\eta_{\textsc{det}})(1-\eta_{\textsc{cav}}^2+2V_{\textsc{x}}\eta_{\textsc{cav}}^3\chi^2)\\
    &+&\rme^{4r}\eta_{\textsc{det}}\left\{\eta_{\textsc{det}}+\eta_{\textsc{cav}}^2\left[2+4V_{\textsc{x}}\eta_{\textsc{cav}}\chi^2-\eta_{\textsc{det}}(4-\eta_{\textsc{cav}}^2(1-4V_{\textsc{x}}\chi^2(1-V_{\textsc{x}}\eta_{\textsc{cav}}\chi^2)))\right]
    \right\}\\
    &+&\rme^{2r}\eta_{\textsc{cav}}^2\eta_{\textsc{det}}^2\left[2-2\eta_{\textsc{cav}}^2+4V_{\textsc{x}}\eta_{\textsc{cav}}\chi^2+8V_{\textsc{x}}^2\eta_{\textsc{cav}}^4\chi^4
    \right]\\
    &+&\eta_{\textsc{cav}}^4\eta_{\textsc{det}}^2[1+4V_{\textsc{x}}\eta_{\textsc{cav}}\chi^2(1+V_{\textsc{x}}\eta_{\textsc{cav}}\chi^2)] {\bigg\}},
    \end{eqnarray*}
    \begin{eqnarray*}
    J_{b}&=&{\bigg\{}\rme^{8r}(1-\eta_{\textsc{det}})^2+2\rme^{6r}\eta_{\textsc{det}}(1-\eta_{\textsc{det}})(1-\eta_{\textsc{cav}}^2+V_{\textsc{x}}\eta_{\textsc{cav}}^3)\\
    &+&\rme^{4r}\eta_{\textsc{det}}\left\{\eta_{\textsc{det}}+\eta_{\textsc{cav}}^2\left[2-(4-\eta_{\textsc{cav}}^2)\eta_{\textsc{det}}+2V_{\textsc{x}}\eta_{\textsc{cav}}(2-(1+\eta_{\textsc{cav}}^2)\eta_{\textsc{det}})\chi^2
    \right]
    \right\}\\
    &+&\rme^{2r}\eta_{\textsc{cav}}^2\eta_{\textsc{det}}^2\left\{2+\eta_{\textsc{cav}}\left[4V_{\textsc{x}}\chi^2-2\eta_{\textsc{cav}}(1+V_{\textsc{x}}\eta_{\textsc{cav}}\chi^2(1-2V_{\textsc{x}}\eta_{\textsc{cav}}\chi^2))
    \right]
    \right\}\\
     &+&\eta_{\textsc{cav}}^4\eta_{\textsc{det}}^2\left[1+4V_{\textsc{x}}\eta_{\textsc{cav}}\chi^2(1+V_{\textsc{x}}\eta_{\textsc{cav}}\chi^2)
    \right]
    {\bigg\}}\bigg/
    {\bigg\{}\rme^{8r}(1-\eta_{\textsc{det}})^2\\
    &+&2\rme^{6r}\eta_{\textsc{det}}(1-\eta_{\textsc{det}})(1-\eta_{\textsc{cav}}^2+2V_{\textsc{x}}\eta_{\textsc{cav}}^3)\\
    &+&\rme^{4r}\eta_{\textsc{det}}\left\{\eta_{\textsc{det}}+\eta_{\textsc{cav}}^2\left[2+4V_{\textsc{x}}\eta_{\textsc{cav}}\chi^2-\eta_{\textsc{det}}(4-\eta_{\textsc{cav}}^2(1-4V_{\textsc{x}}\eta_{\textsc{cav}}\chi^2(1-V_{\textsc{x}}\eta_{\textsc{cav}}\chi^2)))
    \right]
    \right\}\\
    %%%%%%
    &+&\rme^{2r}\eta_{\textsc{cav}}^2\eta_{\textsc{det}}^2\left[2-2\eta_{\textsc{cav}}^2+4V_{\textsc{x}}\eta_{\textsc{cav}}\chi^2+8V_{\textsc{x}}^2\eta_{\textsc{cav}}^4\chi^4
    \right]\\
     &+&\eta_{\textsc{cav}}^4\eta_{\textsc{det}}^2\left[1+4V_{\textsc{x}}\eta_{\textsc{cav}}\chi^2(1+V_{\textsc{x}}\eta_{\textsc{cav}}\chi^2)
    \right]
    {\bigg\}},
    \end{eqnarray*}
    \begin{eqnarray*}
    J_{c}&=&\dfrac{4}{\rme^{5r}(1-\eta_{\textsc{det}})+\rme^{3r}\eta_{\textsc{det}}\left[1-\eta_{\textsc{cav}}^2(1-2V_{\textsc{x}}\eta_{\textsc{cav}}\chi^2)\right]+\rme^{r}\eta_{\textsc{cav}}^2\eta_{\textsc{det}}(1+2V_{\textsc{x}}\eta_{\textsc{cav}}\chi^2)}.
\end{eqnarray*}
For $r=0$, these functions simplify to the form
\begin{eqnarray*}
    J_{a}&=&\frac{1}{2}\left[1+\dfrac{1+4V_{\textsc{x}}\eta_{\textsc{cav}}^3\eta_{\textsc{det}}\chi^2}{1+8V_{\textsc{x}}\eta_{\textsc{cav}}^3\eta_{\textsc{det}}\chi^2(1+2V_{\textsc{x}}\eta_{\textsc{cav}}^3\eta_{\textsc{det}}\chi^2)}
\right]\\
    J_{b}&=&\frac{1}{2}\left[1+\dfrac{1+4V_{\textsc{x}}\eta_{\textsc{cav}}^3\eta_{\textsc{det}}\chi^2}{1+8V_{\textsc{x}}\eta_{\textsc{cav}}^3\eta_{\textsc{det}}\chi^2(1+2V_{\textsc{x}}\eta_{\textsc{cav}}^3\eta_{\textsc{det}}\chi^2)}
\right]\\
    J_{c}&=&\dfrac{4}{1+4V_{\textsc{x}}\eta_{\textsc{cav}}^3\eta_{\textsc{det}}\chi^2}.
\end{eqnarray*}

\section{Verification procedure and inverse maps}\label{verfappendix}

\subsection{Outline of the verification procedure}
 The covariance matrix $\sigma(t)$ of the entangled state after a time $t$ since entanglement generation is given by
\begin{equation}
    \sigma(t)=G^{1/2}\sigma(0)G^{1/2}+(\mathds{1}-G)\sigma_{\textsc{env}},
\end{equation}
and we define the reconstructed covariance matrix $\sigma_{\textsc{ver}}$ so that in the absence of loss $\sigma_{\textsc{ver}}=\sigma(0)$. As in the previous section, we work in a rotating frame at the mechanical frequency only for computational convenience---and hence we don't consider free mechanical evolution in $\sigma(t)$---but this in principle is not necessary. We now give two examples of elements defined for $\sigma_{\textsc{ver}}$ in the optical-mechanical entanglement scheme. The elements of $\sigma_{\textsc{ver}}$ in the mechanical entanglement protocols are defined in precisely the same way.

\textit{$B_{\textsc{ver},12}$ and $C_{\textsc{ver},11}$ in the  optical-mechanical entanglement scheme}--- From Eq.~\eqref{verB}, we have that 
\begin{equation}
    B_{\textsc{ver},12}=\frac{1}{2\chi^2}\left[\mathrm{Var}(\Pv(\frac{\pi}{4}))-\mathrm{Var}(\Pv(\frac{3\pi}{4}))\right],
\end{equation}
and by using $\Pv(\theta)=P_{\textsc{in}}+\chi\XM(\theta)$ one finds that, in the absence of decoherence, 
\begin{eqnarray}
    B_{\textsc{ver},12}&=&\frac{1}{2\chi^2}[\mathrm{Var}(P_{\textsc{in}}+\frac{\chi}{\sqrt{2}}(\XM+\PM))-\mathrm{Var}(P_{\textsc{in}}+\frac{\chi}{\sqrt{2}}(\XM-\PM))]\nonumber\\
    &=&\frac{1}{2}\braket{\{\XM,\PM\}}-\braket{\XM}\braket{\PM}\nonumber\\\
    &=&B_{12}(0).
\end{eqnarray}
Similarly, from Eq.~\eqref{verC} we have that 
\begin{equation}
   C_{\textsc{ver},11}=\frac{1}{2\chi}\left[ \mathrm{Var}(\Pv''(2\pi))-\mathrm{Var}(\XL''(2\pi))\right].
\end{equation}
$\Pv''(\theta)$ and $\XL''(\theta)$ are defined implicitly in the main text and are given by $\Pv''(\theta)=\frac{1}{\sqrt{2}}(\Pv(\theta)+\XL)$ and $\XL''(\theta)=\frac{1}{\sqrt{2}}(\XL-\Pv(\theta))$, respectively. Therefore, we have
\begin{eqnarray}
    C_{\textsc{ver},11}&=&\frac{1}{4\chi}\left[\mathrm{Var}(\Pv(2\pi)+\XL)-\mathrm{Var}(\XL-\Pv(2\pi))\right]\nonumber\\
    &=&\frac{1}{2}\braket{\{\XL,\XM\}}-\braket{\XL}\braket{\XM}\nonumber\\\
    &=&C_{11}(0).
\end{eqnarray}

However, the reconstruction process will be imperfect due to decoherence processes and as such $\sigma_{\textsc{ver}}\neq\sigma(0)$. 
A \textit{conservative in time} approach is therefore taken to ensure that entanglement is not overestimated and a \textit{conservative in time and noise} approach is also taken when $\mathrm{Var}(P_{\textsc{in}})$ is unknown and cannot be subtracted.

To describe the effect of decoherence in the reconstruction process and address the fact that all the matrix elements of $\sigma_{\textsc{ver}}$ are accessed at different times, we use the elements of $\sigma(t)$ to calculate those of $\sigma_{\textsc{ver}}$. Describing the elements of $\sigma_{\textsc{ver}}$ using those of $\sigma(t)$ is slightly more complicated when the elements of $\sigma_{\textsc{ver}}$ depend on two different mechanical phase-space angles $\theta$---and therefore two different times $t=\theta/\omega_{\textsc{m}}$, as illustrated in the calculation of $B_{\textsc{ver},12}$ below. 

\textit{$B_{\textsc{ver},22}$, an element that depends on one time}--- The $B_{\textsc{ver},22}$ element is given by
\begin{eqnarray}
B_{\textsc{ver},22}&=&\frac{1}{\chi^2}\left[\mathrm{Var}(\Pv(\frac{\pi}{2}))-\mathrm{Var}(P_{\textsc{in}})\right]\nonumber\\
&=&\left\{\begin{array}{c} 
B_{22}(\frac{\pi}{2}) \quad \mbox{conservative in time}\\
B_{22}(\frac{\pi}{2})+\frac{1}{2\chi^2} \quad \mbox{conservative in time and noise}.
\end{array}
\right.
\end{eqnarray}
The conservative in time and noise approach corresponds to not subtracting $\mathrm{Var}(P_{\textsc{in}})$ and here we assume that vacuum noise contaminates $B_{\textsc{ver},22}$. 

\textit{$B_{\textsc{ver},12}$, an element that depends on two times}--- As above
\begin{gather}
B_{\textsc{ver},12}=\frac{1}{2\chi^2}\left[\mathrm{Var}(\Pv(\frac{\pi}{4}))-\mathrm{Var}(\Pv(\frac{3\pi}{4}))\right]\nonumber\\
\hspace{-0.5cm}=\frac{1}{4}\left[B_{11}(\frac{\pi}{4})-B_{11}(\frac{3\pi}{4})+B_{22}(\frac{\pi}{4})-B_{22}(\frac{3\pi}{4})+2B_{12}(\frac{\pi}{4})+2B_{12}(\frac{3\pi}{4})\right].
\end{gather}
In the absence of decoherence, $\sigma(0)\equiv\sigma(t)$ and so $B_{\textsc{ver},12}=B_{12}(0)$.

\subsection{Including optical loss in $\sigma_{\textsc{ver}}$}

If we include optical losses in the verification stage, the formulas for $\sigma_{\textsc{ver}}$ in the main text are modified. To derive these new forms for $\sigma_{\textsc{ver}}$, we assume that the variance of each loss mode is equal to that of $P_{\textsc{in}}$ and at each loss channel an optical quadrature $Q$ transforms according to
\begin{equation}
Q\rightarrow\sqrt{\eta}Q+\sqrt{1-\eta}Q_{\textsc{vac}},
\end{equation}
where $Q_{\textsc{vac}}$ is the environmental quadrature for a given channel and $\mathrm{Var}(Q_{\textsc{vac}})=\mathrm{Var}(P_{\textsc{in}})$. Below we present how the formulas for $\sigma_{\textsc{ver}}$ are modified in each entanglement scheme.

\textit{Optical-mechanical entanglement scheme---} By including optical losses in the verification stage, the $A_{\textsc{ver}}$ block is modified as follows:
\begin{equation*}
A_{\textsc{ver}}\rightarrow\frac{1}{\eta}\begin{pmatrix}\mathrm{Var}(\XL)-(1-\eta)\mathrm{Var}(P_{\textsc{in}})&\frac{1}{2}[\mathrm{Var}(\PL(\frac{3\pi}{4}))-\mathrm{Var}(\PL(\frac{\pi}{4}))]\\\frac{1}{2}[\mathrm{Var}(\PL(\frac{3\pi}{4}))-\mathrm{Var}(\PL(\frac{\pi}{4}))]&\mathrm{Var}(\PL)-(1-\eta)\mathrm{Var}(P_{\textsc{in}})\end{pmatrix}.
\end{equation*}
The $B_{\textsc{ver}}$ and $C_{\textsc{ver}}$ blocks rescale according to $B_{\textsc{ver}}\rightarrow B_{\textsc{ver}}/\eta$,
and $C_{\textsc{ver}}\rightarrow\ C_{\textsc{ver}}/\eta$.

\textit{Interferometric mechanical entanglement scheme---} In this scheme, all the elements of the reconstructed covariance matrix rescale in the same way $\sigma_{\textsc{ver}}\rightarrow\sigma_{\textsc{ver}}/\eta$.

\textit{Non-interferometric mechanical entanglement scheme---} Here, the $A_{\textsc{ver}}$ and $B_{\textsc{ver}}$ blocks are modified in the same way $A_{\textsc{ver}}\rightarrow A_{\textsc{ver}}/\eta$ and $B_{\textsc{ver}}\rightarrow B_{\textsc{ver}}/\eta$. While $C_{\textsc{ver}}$ is modified to $C_{\textsc{ver}}\rightarrow C_{\textsc{ver}}/\eta^2$.

Assuming that $\eta$ is a well-known quantity, none of the results in the reconstruction stage are affected.

\subsection{Inverse map}
Once $\sigma_{\textsc{ver}}$ has been reconstructed, an inverse map may be applied to calculate the covariance matrix $\sigma(0)$ of the bipartite state at the moment of entanglement generation. In the verification procedure, the elements of $\sigma_{\textsc{ver}}$ are defined in terms of the state $\sigma(t)$. From Eq.~\eqref{covdecomap}, for the elements of $\sigma_{\textsc{ver}}$ defined at a single time, we have
\begin{eqnarray}
\sigma_{\textsc{ver},ij}&=&\sigma_{ij}(t_{ij})\nonumber\\
&=&\left[G^{1/2}\sigma(0)G^{1/2}+(\mathds{1} - G)\sigma_{\textsc{env}}\right]_{ij}.
\end{eqnarray}
Here, each element of $\sigma_{\textsc{ver}}$ may be defined at different times, hence the inclusion of the indices $ij$ to the temporal parameter $t_{ij}$. Using the fact that $G$ and $\sigma_{\textsc{env}}$ are diagonal matrices we may simplify the expression for $\sigma_{\textsc{ver},ij}$ to
\begin{eqnarray}
\sigma_{\textsc{ver},ij}=G^{1/2}_{ii}\sigma(0)_{ij}G^{1/2}_{jj}+(\mathds{1} - G)_{ii}\sigma_{\textsc{env},jj}\delta_{ij},
\end{eqnarray}
or inversely
\begin{equation}\label{onetime}
\sigma(0)_{ij}=G^{-1/2}_{ii}\left[
\sigma_{\textsc{ver},ij}-(\mathds{1} - G)_{ii}\sigma_{\textsc{env},jj}\delta_{ij}\right]G^{-1/2}_{jj}.
\end{equation}

Elements defined at two different times, $t_{ij,1}$ and $t_{ij,2}$, may be written as
\begin{eqnarray}
\sigma_{\textsc{ver},ij}&=&\sum_{k}\sum_{nm}c_{nm,k}\sigma_{nm}(t_{ij,k})\nonumber\\
&=&\sum_{k}\sum_{nm\neq{ij}}c_{nm,k}\sigma_{nm}(t_{ij,k})+\sum_{k}c_{ij,k}\sigma_{ij}(t_{ij,k}).
\end{eqnarray}
Here, $c_{nm,k}$ are real coefficients and $\sigma_{ij}(t_{ij,k})$ is given by
\begin{equation}\label{onetwotime}
    \sigma_{ij}(t_{ij,k})=G^{1/2}_{ii}(t_{ij,k})\sigma(0)_{ij}G^{1/2}_{jj}(t_{ij,k})+(\mathds{1} - G)_{ii}(t_{ij,k})\sigma_{\textsc{env},jj}\delta_{ij},
\end{equation}
where a temporal argument has to now be added to the matrix $G$ to facilitate the two different times. Rearranging the expression for $\sigma_{\textsc{ver},ij}$ leads to 
\begin{eqnarray}\label{twotime}
\hspace{-1cm}\sigma(0)_{ij}=\dfrac{\sigma_{\textsc{ver},ij}-\sum_{k}\sum_{nm\neq{ij}}c_{nm,k}\sigma_{nm}(t_{ij,k})-\sum_{k}c_{ij,k}(\mathds{1} - G)_{ii}(t_{ij,k})\sigma_{\textsc{env},jj}\delta_{ij}}{\sum_{k}c_{ij,k}G^{1/2}_{ii}(t_{ij,k})G^{1/2}_{jj}(t_{ij,k})}.\hspace{0.5cm}
\end{eqnarray}
Providing the quantities $\sigma_{nm}(0)$, with $nm\neq{ij}$, have been calculated from the elements of $\sigma_{\textsc{ver}}$ defined at one time, then the quantities $\sigma_{nm}(t_{ij,k})$ may also be calculated. After this, the elements $\sigma(0)_{ij}$ can be calculated from the elements of $\sigma_{\textsc{ver}}$ that depend on two times. To summarise: in order to calculate $\sigma(0)$, the inverse map is applied to elements of $\sigma_{\textsc{ver}}$ that depend on one time and these are then used to find the elements which depend on two times. Below is a short example of how this procedure works for the $B_{\textsc{ver}}$ block of the optical-mechanical covariance matrix.

\textit{Calculating elements of $\sigma(0)$ using the inverse map}--- Consider again the element $B_{\textsc{ver},12}$ given by
\begin{equation}
    B_{\textsc{ver},12}=\frac{1}{4}\left[B_{11}(\frac{\pi}{4})-B_{11}(\frac{3\pi}{4})+B_{22}(\frac{\pi}{4})-B_{22}(\frac{3\pi}{4})+2B_{12}(\frac{\pi}{4})+2B_{12}(\frac{3\pi}{4})
    \right].
\end{equation}
The elements $B_{\textsc{ver},11}=B_{11}(0)$ and $B_{\textsc{ver},22}=B_{22}(\frac{\pi}{2})$ only depend on one time, and so we have that $B_{11}(0)$ and $B_{22}(0)$ are given by Eq.~\eqref{onetime}. Explicitly we have 
\begin{eqnarray}
B_{11}(0)&=&B_{\textsc{ver},11},\\
B_{22}(0)&=&G^{-1/2}_{22}(\pi/2)\left[B_{\textsc{ver},22}-(\mathds{1}-G)_{22}\sigma_{\textsc{env},22}\right]G^{-1/2}_{22}(\pi/2).
\end{eqnarray}
This allows one to work out $B_{11}(\frac{\pi}{4})$, $B_{11}(\frac{3\pi}{4})$,$B_{22}(\frac{\pi}{4})$, and $B_{22}(\frac{3\pi}{4})$ using Eq.~\eqref{onetwotime}. Eq.~\eqref{twotime} may then be used to find an expression for $B_{12}(0)$. Namely, 
\begin{equation}
   B_{12}(0)=\dfrac{B_{\textsc{ver},12}-\frac{1}{4}\left[B_{11}(\frac{\pi}{4})-B_{11}(\frac{3\pi}{4})+B_{22}(\frac{\pi}{4})-B_{22}(\frac{3\pi}{4})
    \right]}
    {\frac{1}{2}G_{11}^{1/2}(\pi/4)G_{11}^{1/2}(\pi/4)+\frac{1}{2}G_{11}^{1/2}(3\pi/4)G_{11}^{1/2}(3\pi/4)}.
\end{equation}

The above analysis assumes that the inverse maps can be applied perfectly, whereas in reality this will not be the case. Uncertainties in the values of $\eta$ and $\gamma$ will limit the precision to which these inverse operations can be applied and hence limit the amount of entanglement one can verify when reconstructing $\sigma(0)$ from $\sigma_{\textsc{ver}}$.

\section{Single mode Gaussian state of fixed purity in a phase-insensitive decoherence channel}\label{channel}
Consider the covariance matrix
\begin{equation}
    \sigma=\begin{pmatrix}a\rme^{2r}&0\\0&b\rme^{-2r}
    \end{pmatrix},
\end{equation}
which characterizes a general single mode Gaussian state of fixed purity $\mu(\sigma)=1/\sqrt{4\mathrm{Det}\sigma}=1/\sqrt{4ab}$ up to a local rotation. After the state passes through a phase-insensitive decoherence channel $\mathcal{E}_{\textsc{g}}$, with $G=\eta\mathds{1}_{2}$ and $\sigma_{\mathrm{env}}$ corresponding to a thermal state with occupation $\bar{N}$, the covariance matrix is given by
\begin{equation}
    \sigma'=\eta\begin{pmatrix}a\rme^{2r}+\delta&0\\0&b\rme^{-2r}+\delta
    \end{pmatrix}.
\end{equation}
Here, $\delta=\frac{(1-\eta)(1+2\bar{N})}{2\eta}>0$. Maximizing $\mu(\sigma')$ with respect to $r$ gives $\rme^{2r}=\sqrt{b/a}$ and hence the input state $\sigma$ which maximizes the output purity from the channel is $\sigma=\sqrt{ab}\mathds{1}_{2}$. 

Likewise, the von-Neumann entropy of the output state~\cite{olivares2012} is given by
\begin{equation}
    s_{\textsc{v}}(\sigma')=(\nu+\frac{1}{2})\log_{2}\left(\nu+\frac{1}{2}\right)-(\nu-\frac{1}{2})\log_{2}\left(\nu-\frac{1}{2}\right)
\end{equation}
where $\nu$ is the symplectic eigenvalue of $\sigma'$---given by
\begin{equation}
    \nu=\eta\sqrt{(a\rme^{2r}+\delta)(b\rme^{-2r}+\delta)}.
\end{equation}
The output entropy is minimized with $\rme^{2r}=\sqrt{b/a}$, which again leads to a symmetric input state.

\section{Avoiding non-monotonic behaviour}\label{nonmonappendix}

As mentioned in Section \ref{nonmoncons}, when the squeezing operation can be applied before the optical decoherence channel at the cavity output, the non-monotonic behaviour in logarithmic negativity in the optical-mechanical and interferometric entanglement schemes is avoided. This is because the entangled state is protected from loss of information to the environment before it enters the decoherence channels. In this case, as the optical loss channels are now consecutive, and occur after the optical squeezing operations, they may be combined into a single loss channel with efficiency $\eta=\eta_{\textsc{cav}}\eta_{\textsc{det}}$. Below, we present a brief study of how optical squeezers are used in the optical-mechanical entanglement scheme to avoid non-monotonic behaviour in this case. 

The PPT entanglement criterion reads
\begin{eqnarray}
\tilde{\nu}_{-}=\begin{cases}
\geq\frac{1}{2}&\mathrm{separable}\\
<\frac{1}{2}&\mathrm{entangled},
\end{cases}
\end{eqnarray}
and can be rewritten as
\begin{eqnarray}\label{lambda}
\Lambda=\begin{cases}
\geq0&\mathrm{separable}\\
<0&\mathrm{entangled}
\end{cases}.
\end{eqnarray}
Here,
\begin{eqnarray}
\Lambda=4\mathrm{Det}\sigma-\tilde{\Delta}+\frac{1}{4}
\end{eqnarray}
and $\sigma$ refers to the optomechanical state formed by Eq.~\eqref{omintsq} when the order of the $U_{\textsc{sq}}$ and $\mathcal{E}_{\eta_{\textsc{cav}}}$ operations is swapped.
We may expand $\Lambda$ in the parameter $\chi$ to study the PPT criterion in the limit $\chi\to\infty$. For brevity, this expansion is not reported here, but we find that, with an ansatz for the optical squeezing parameter of the form $r=\ln(1+c\chi^2)$, it is possible to ensure that $\lim_{\chi\to\infty}\Lambda<0$ if $V_{\textsc{p}}<f(c,\bar{N},\eta,\rme^{\gamma t})$. Here, $f$ is a function of the system parameters and represents the minimal cooling requirement to observe entanglement in the large $\chi$ limit---hence avoiding non-monotonic behaviour. Maximising the upper bound over all $c$ gives the cooling requirement for entanglement to persist as $\chi\to\infty$ 
\begin{eqnarray}
V_{\textsc{p}}&<&\frac{2\left[f_{1}(\eta,\rme^{\gamma t},\bar{N})+f_{2}(\eta,\rme^{\gamma t},\bar{N})\right]}{
    \rme^{-\gamma t}(1+2\bar{N})
(f_{3}(\eta,\rme^{\gamma t},\bar{N}))},\\
f_{1}(\eta,\rme^{\gamma t},\bar{N})&=&\bar{N}(1+\bar{N})[\rme^{-4\gamma t}-(3-\eta)\rme^{-3\gamma t}+3(1-\eta)\rme^{-2\gamma t}+(3\eta-1)\rme^{-\gamma t}-\eta],\nonumber\\
f_{2}(\eta,\rme^{\gamma t},\bar{N})&=&-\rme^{-3\gamma t}+(1-\eta)\rme^{-2\gamma t}+\rme^{-\gamma t}\eta,\nonumber\\
f_{3}(\eta,\rme^{\gamma t},\bar{N})&=&\rme^{-3\gamma t}-(2-\eta)\rme^{-2\gamma t}+(1-2\eta)\rme^{-\gamma t}+\eta\nonumber.
\end{eqnarray}
The $c$ which maxmizes $f$ is given by
\begin{equation}
c=(1+2\bar{N})\frac{1-\rme^{-\gamma t}}{1+\rme^{-\gamma t}}.
\end{equation}
For a genuine mechanical covariance matrix we also require $V_{\textsc{p}}>0$. When decoherence processes dominate, $f$ becomes negative and it is no longer possible to avoid separability in the large $\chi$ limit. We also note that this cooling condition does not depend on the position variance of the mechanical precooled state $V_{\textsc{x}}$. This is because the optomechanical unitary generates phase-space displacements in an orthogonal direction to the position quadrature.

\section{Krauss operator approach} \label{MeasOpApproach}
In this appendix, we derive the Krauss operators that describe the entanglement protocols of Sections \ref{light-mechanics} and \ref{twomechanics}. This Krauss operator representation allows one to further understand the dependence of logarithmic negativity on homodyne measurement angles, and the balance between unitary and positive operations in the generation of entanglement.

As this approach will be completely equivalent to the symplectic formalism outlined in the main text, and so as not to obscure the key elements of this discussion, we neglect the effects of optical loss and mechanical decoherence. However, if one wishes, these deleterious effects may be included: each optical or mechanical loss channel introduces another Hilbert space one would need to consider when calculating the Krauss operator, and hence introduces an index to the Krauss operator indicating the number of photons or phonons which leak to the environment. A sum over all possible photon and phonon numbers detected by the environment would then be performed to describe the lossy Krauss operation. In what follows we take the amplitude of the coherent pulse to be real without loss of generality, $\alpha=X_{\alpha}/\sqrt{2}\in\mathds{R}$.

\subsection{Optical-mechanical and the precooling stage}

The Krauss operator corresponding to the optical-mechanical entanglement protocol is given simply by
\begin{equation}
\Upsilon(\XL,\XM)=U_{\textsc{sq}}(r)\rme^{\rmi\chi\XL\XM}.
\end{equation}
The two mode unitary $\rme^{\rmi\chi\XL\XM}$ entangles the optical and mechanical states, while the optical squeezer protects the state from optical loss and mechanical decoherence.

$\Upsilon(\XL,\XM)$ is completely unitary, which reflects the continuous behaviour of the entanglement montone with respect to the system parameters, such as  interaction strength $\chi$ and optical efficiency $\eta$.

The Krauss operator $\Upsilon(\XM,\varphi)$ describing the precooling stage results from a single optomechanical interaction, where the input light is in the state $\ket{\alpha}$, followed by optical squeezing and homodyne detection at angle $\varphi$. Hence,
\begin{gather}
\Upsilon(\XM,\varphi)=\bra{\XL({\varphi})}U_{\textsc{sq}}(r)\rme^{\rmi\chi\XL\XM}\ket{\alpha}\nonumber\\
=\rme^{\rmi\chi{X}_{\alpha}\XM/2}\bra{\XL(\varphi)}U_{\textsc{sq}}(r)\ket{(X_{\alpha}+\rmi\chi\XM)/\sqrt{2}}.
\end{gather}
Therefore, 
\begin{gather}
\Upsilon(\XM,\varphi)=(\Gamma_{\textsc{r}}/\pi)^{1/4}\exp\{-\Gamma(\XL(\varphi)-X_{0})^2/2\}\nonumber\\
\exp\{\rmi{P}_{0}\XL(\varphi)\}
\exp{-\rmi{X_{0}P_{0}}/2}\exp\{\rmi\chi{X}_{\alpha}\XM/2\}.
\end{gather}
Here,
\begin{gather}
    X_{0}=\rme^{r}X_{\alpha}\cos\varphi+\rme^{-r}\chi\XM\sin\varphi,\\
    P_{0}=\rme^{-r}\chi\XM\cos\varphi-\rme^{r}X_{\alpha}\sin\varphi,\\
    \Gamma(\varphi)=\Gamma_{\textsc{r}}(\varphi)+\rmi\Gamma_{\textsc{i}}(\varphi),\\
    \Gamma_{\textsc{r}}(\varphi)=\frac{1}{\cosh2r+\cos2\varphi\sinh2r},\\
    \Gamma_{\textsc{i}}(\varphi)=\frac{2\sin2\varphi\sinh2r}{\cosh2r+\cos2\varphi\sinh2r}.
\end{gather}

The optimal homodyne angle to precool the mechanical mode via measurement is $\varphi=\pi/2$. This is because the mechanical position quadrature is imprinted on the optical phase quadrature of light. In this case the Krauss operator is
\begin{equation}
  \Upsilon(\XM,\pi/2)=(\rme^{2r}/\pi)^{1/4}\exp\{-\rme^{2r}(\PL-\rme^{-r}\chi\XM)^2/2-\rmi\rme^{r}{X}_{\alpha}\PL\}.
\end{equation}
We write the action of the Krauss operator on the mechanical mode as $\Upsilon\circ\rho$. Applying the Krauss operator $m$ times produces the state $\Upsilon^{m}\circ\rho$, where
\begin{gather}
[\Upsilon(\XM,\varphi)]^m=(\Gamma_{\textsc{r}}/\pi)^{m/4}\exp\{-\Gamma(\sqrt{m}\XL(\varphi)-\sqrt{m}X_{0})^2/2\}\nonumber\\
\exp\{\rmi{m}{P}_{0}\XL(\varphi)\}
\exp{-\rmi{m}{X_{0}P_{0}}/2}\exp\{\rmi{m}\chi{X}_{\alpha}\XM/2\}.
\end{gather}
This operation is equivalent to a single application of the Krauss operator but with rescalings $\chi\rightarrow\sqrt{m}\chi$, $\alpha\rightarrow\sqrt{m}\alpha$, and $\XL(\varphi)\rightarrow\sqrt{m}\XL(\varphi)$. Hence, increasing the number of pulsed optomechanical interactions by a factor $m$ is equivalent to a single pulsed interaction with $m$-times more photons.

\subsection{Interferometric mechanical entanglement}
The Krauss operator which acts on the initially separable state of the two mechanical oscillators, corresponding to Eq.~\eqref{DMMap}, is
\begin{gather}
\Upsilon(\XMo,\XMt,\varphi,\psi)=\bra{\XLo({\varphi})}\bra{\XLt({\psi})}U_{\textsc{bs}}U_{\textsc{sq}}^{\otimes2}\rme^{\rmi\chi\XLt\XMt}\rme^{\rmi\chi\XLo\XMo}U_{\textsc{bs}}\ket{\sqrt{2}\alpha}\ket{0}
\end{gather}
Up to irrelevant phase terms, the Krauss operator is therefore given by
\begin{gather}
\Upsilon(\XMo,\XMt,\varphi,\psi)=(\Gamma_{\textsc{r}}(\varphi)\Gamma_{\textsc{r}}(\psi)/\pi^2)^{1/4}\exp\{-\Gamma(\varphi)(\XLo(\varphi)-X_{1})^2/2\}
\exp\{\rmi{P}_{1}\XLo(\varphi)\}\nonumber\\
\exp{-\rmi{X_{1}P_{1}}/2}
\exp\{-\Gamma(\psi)(\XLt(\psi)-X_{2})^2/2\}
\exp\{\rmi{P}_{2}\XLt(\psi)\}
\exp{-\rmi{X_{2}P_{2}}/2},
\end{gather}
where $\Gamma$ is defined as above, while 
\begin{gather}
    X_{1}=\rme^{-r}\chi(\XMo+\XMt)\sin\varphi/\sqrt{2},\\
   P_{1}=\rme^{-r}\chi(\XMo+\XMt)\cos\varphi/\sqrt{2},\\
    X_{2}=-\sqrt{2}\rme^{r}X_{\alpha}\cos\psi-\rme^{-r}\chi(\XMo-\XMt)\sin\psi/\sqrt{2},\\
    P_{2}=\sqrt{2}\rme^{r}X_{\alpha}\sin\psi-\rme^{-r}\chi(\XMo-\XMt)\cos\psi/\sqrt{2}.
\end{gather}

In this case, the polar decomposition of the Krauss operator is given by $\Upsilon(\XMo,\XMt,\varphi,\psi)=U(\XMo,\XMt,\varphi,\psi)P(\XMo,\XMt,\varphi,\psi)$, with positive operator 
\begin{gather}
    P(\XMo,\XMt,\varphi,\psi)=(\Gamma_{\textsc{r}}(\varphi)\Gamma_{\textsc{r}}(\psi)/\pi^2)^{1/4}\exp\{-\Gamma_{\textsc{r}}(\varphi)(\XLo(\varphi)-X_{1})^2/2\}\nonumber\\
\exp\{-\Gamma_{\textsc{r}}(\psi)(\XLt(\psi)-X_{2})^2/2\},
\end{gather}
and effective unitary operator
\begin{gather}
    U(\XMo,\XMt,\varphi,\psi)=(\Gamma_{\textsc{r}}(\varphi)\Gamma_{\textsc{r}}(\psi)/\pi^2)^{1/4}\exp\{-\rmi\Gamma_{\textsc{i}}(\varphi)(\XLo(\varphi)-X_{1})^2/2\}
\exp\{\rmi{P}_{1}\XLo(\varphi)\}\nonumber\\
\exp{-\rmi{X_{1}P_{1}}/2}
\exp\{-\rmi\Gamma_{\textsc{i}}(\psi)(\XLt(\psi)-X_{2})^2/2\}
\exp\{\rmi{P}_{2}\XLt(\psi)\}
\exp{-\rmi{X_{2}P_{2}}/2},
\end{gather}
which also depends on the homodyne measurement outcomes. The parts of these operators which depend on the square of the mechanical EPR quadratures may induce entanglement as they contain terms proportional to the product $\XMo\XMt$. We call these parts $P_{\textsc{ent}}$ and $U_{\textsc{ent}}$, respectively, and they are given by
\begin{gather}
   P_{\textsc{ent}}=(\Gamma_{\textsc{r}}(\varphi)\Gamma_{\textsc{r}}(\psi)/\pi^2)^{1/4}\exp\{-\Gamma_{\textsc{r}}(\varphi)\rme^{-2r}\chi^2(\XMo+\XMt)^2\sin^2\varphi/4\}\nonumber\\
\exp\{-\Gamma_{\textsc{r}}(\psi)\rme^{-2r}\chi^2(\XMo-\XMt)^2\sin^2\psi/4\},
\end{gather}
and 
\begin{gather}
   U_{\textsc{ent}}=(\Gamma_{\textsc{r}}(\varphi)\Gamma_{\textsc{r}}(\psi)/\pi^2)^{1/4}\exp\{-\rmi\Gamma_{\textsc{i}}(\varphi)\rme^{-2r}\chi^2(\XMo+\XMt)^2\sin^2\varphi/4\}\nonumber\\
\exp\{-\rmi\Gamma_{\textsc{i}}(\psi)\rme^{-2r}\chi^2(\XMo-\XMt)^2\sin^2\psi/4\}
\exp{-\rmi\rme^{-2r}\chi^2(\XMo+\XMt)^2\sin\varphi\cos\varphi/4}\nonumber\\
\exp{-\rmi\rme^{-2r}\chi^2(\XMo-\XMt)^2\sin\psi\cos\psi/4}
\end{gather}
As stated in the main text, the homodyne configurations $(\varphi,\psi)=(0,\pi/2)$ and $(\varphi,\psi)=(\pi/2,0)$ therefore corresponds to Bayesian inference of the mechanical EPR quadratures $(\XMo+\XMt)$ and $(\XMo-\XMt)$, respectively. To see this explicitly, consider the case when $(\varphi,\psi)=(0,\pi/2)$, then we have that
\begin{gather}
    P_{\textsc{ent}}=\exp{-\chi^2(\XMo-\XMt)^2/4},\nonumber\\
    U_{\textsc{ent}}=\mathds{1}.
\end{gather}
Hence, the total operation $\Upsilon(\XMo,\XMt,\theta,\phi)$ corresponds to Bayesian inference up to local unitaries.

Whilst at $(\varphi,\psi)=(\pi/4,3\pi/4)$ and $(\varphi,\psi)=(3\pi/4,\pi/4)$ the measurement induces a two-mode entangling unitary operation and the positive operator $P_{\textsc{ent}}$ does not lead to any entanglement generation. From a Bayesian perspective, this operation amounts to unitarily evolving the prior towards a more entangled state. Taking $(\varphi,\psi)=(\pi/4,3\pi/4)$, we have that
\begin{gather}
    P_{\textsc{ent}}=\exp{-\rme^{-2r}\chi^2(\XMo^2+\XMt^2)/4\cosh2r},\nonumber\\
    U_{\textsc{ent}}=\exp{-\rmi\rme^{-2r}\chi^2(1+2\tanh2r)\XMo\XMt/2}.
\end{gather}

\begin{figure}
\vspace{-3cm}
     \hspace{-1cm}\includegraphics[width=0.93481\linewidth,angle=270]{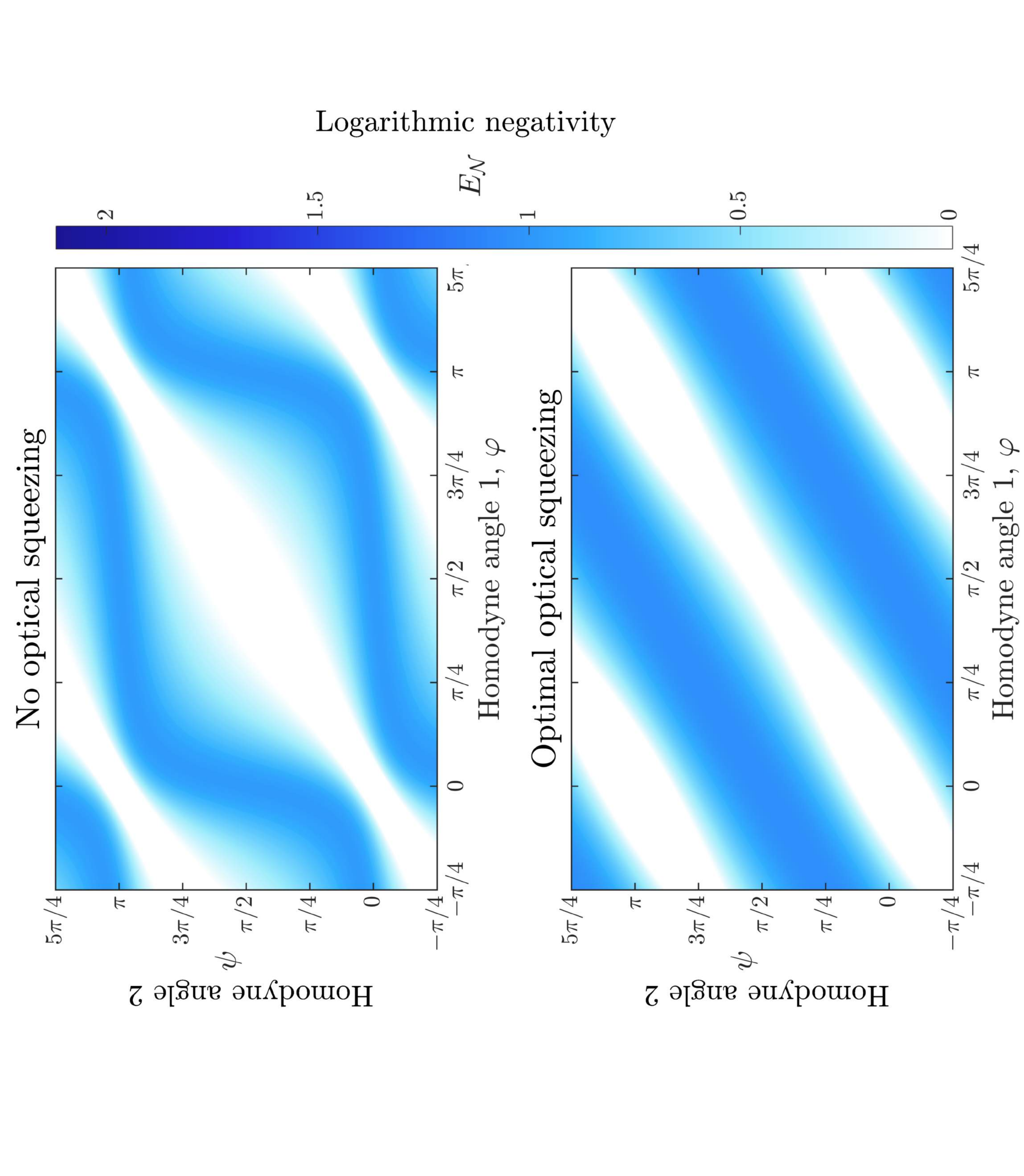}
    \caption{\textbf{Dependence of the logarithmic negativity, with and without optical squeezing, as a function of the optical homodyne angles $\varphi$ and $\psi$ in the interferometric scheme.} Here, $E_{\mathcal{N}}$ is plotted after the mechanical modes have rotated and decohered through a quarter of a period since entanglement generation. The contour plot with no optical squeezing has $\chi=2.97$, while the plot with optimal optical squeezing has $r=0.57$ and $\chi=3.15$. These values are the same as those obtained from Fig.~\ref{intlogneg}, which maximize the logarithmic negativity. Without optical squeezing, the logarithmic negativity reaches a maximum of $E_{\mathcal{N}}=0.98$, while with optimal optical squeezing, $E_{\mathcal{N}}=1.07$ is reached. There are a continuum of maximum in the region $\varphi=[0,\pi)$ and $\psi=[0,\pi)$, and we plot an extra $\pi/4$ outside this range to demonstrate the oscillatory nature more clearly.
    }
    \label{intmechangles}
\end{figure}

Fig.~\ref{intmechangles} shows the balance between Bayesian inference and unitary action with and without the use of optical squeezers. These contour plots for the logarithmic negativity, show that there is a continuum of optimal homodyne configurations, regardless of whether or not optical squeezers are implemented. The configurations $(\varphi,\psi)=(0,\pi/2)$ and $(\varphi,\psi)=(\pi/2,0)$ chosen in the main text correspond to just two of these optimal choices.

When optical squeezers are not used the contour plot is more heavily weighted around the set of points $(\varphi,\psi)=\{(0,\pi/2),(\pi/2,0),(\pi,\pi/2),(\pi/2,\pi)\}$, which corresponds to maximizing the contribution of $P_{\textsc{ent}}$, as compared to the regions surrounding $(\varphi,\psi)=\{(0,\pi/2),(3\pi/4,\pi/4)\}$, which corresponds to maximizing the entangling effect of $U_{\textsc{ent}}$. This observation demonstrates that Bayesian inference of the mechanical EPR quadratures is the more effective way to create mechanical entanglement from the optical-mechanical entanglement. This is because, without the use of squeezers, the optical subspace is asymmetric with more information about the mechanical position quadratures encoded in the phase quadratures of the optical modes. 

It is interesting to note that when the optimal choice of optical squeezers is utilized, the contour plot shows a more equal balance between the effects of $P_{\textsc{ent}}$ and $U_{\textsc{ent}}$. With the best choice of homodyne angles corresponding to approximately $\psi=\varphi-\pi/2$ and $\psi=\varphi+\pi/2$. Optical squeezers bring the optical modes to a more symmetric state, and hence the contour plot is less heavily weighted to areas that correspond to a combination of a phase and an amplitude quadrature measurement.

As mentioned in the main text, no entanglement can be generated when $\varphi=\psi$ as this allows one to gain knowledge of $\XMo$ and $\XMt$ separately.

\subsection{Non-interferometric mechanical entanglement}
The Krauss operator that acts on the initial separable mechanical state, corresponding to Eq.~\ref{nonintmap}, is given by 
\begin{gather}
\Upsilon(\XMo,\XMt,\varphi)=\bra{\XL({\varphi})}U_{\textsc{sq}}\rme^{\rmi\chi\XL\XMt}U_{\textsc{sq}}\rme^{\rmi\chi\XL\XMo}\ket{\alpha}
\end{gather}
Up to irrelevant local unitaries, the Krauss operator may be expressed as
\begin{gather}
\Upsilon(\XMo,\XMt,\varphi)=(\Gamma_{\textsc{r}}/\pi)^{1/4}\exp\{-\Gamma(\XL(\varphi)-X_{0})^2/2\}\nonumber\\
\exp\{\rmi{P}_{0}\XL(\varphi)\}
\exp{-\rmi{X_{0}P_{0}}/2}.
\end{gather}
Here,
\begin{gather}
    X_{0}=\rme^{2r}X_{\alpha}\cos\varphi+\chi(\rme^{-2r}\XMo+\rme^{-r}\XMt)\sin\varphi,\\
    P_{0}=\chi(\rme^{-2r}\XMo+\rme^{-r}\XMt)\cos\varphi-\rme^{2r}X_{\alpha}\sin\varphi,\\
    \Gamma(\varphi)=\Gamma_{\textsc{r}}(\varphi)+\rmi\Gamma_{\textsc{i}}(\varphi),\\
    \Gamma_{\textsc{r}}(\varphi)=\frac{1}{\cosh4r+\cos2\varphi\sinh4r},\\
    \Gamma_{\textsc{i}}(\varphi)=\frac{2\sin2\varphi\sinh4r}{\cosh4r+\cos2\varphi\sinh4r}.
\end{gather}
\begin{figure}
\vspace{-3cm}
    \centering
    \includegraphics[width=0.4572\linewidth,angle=270]{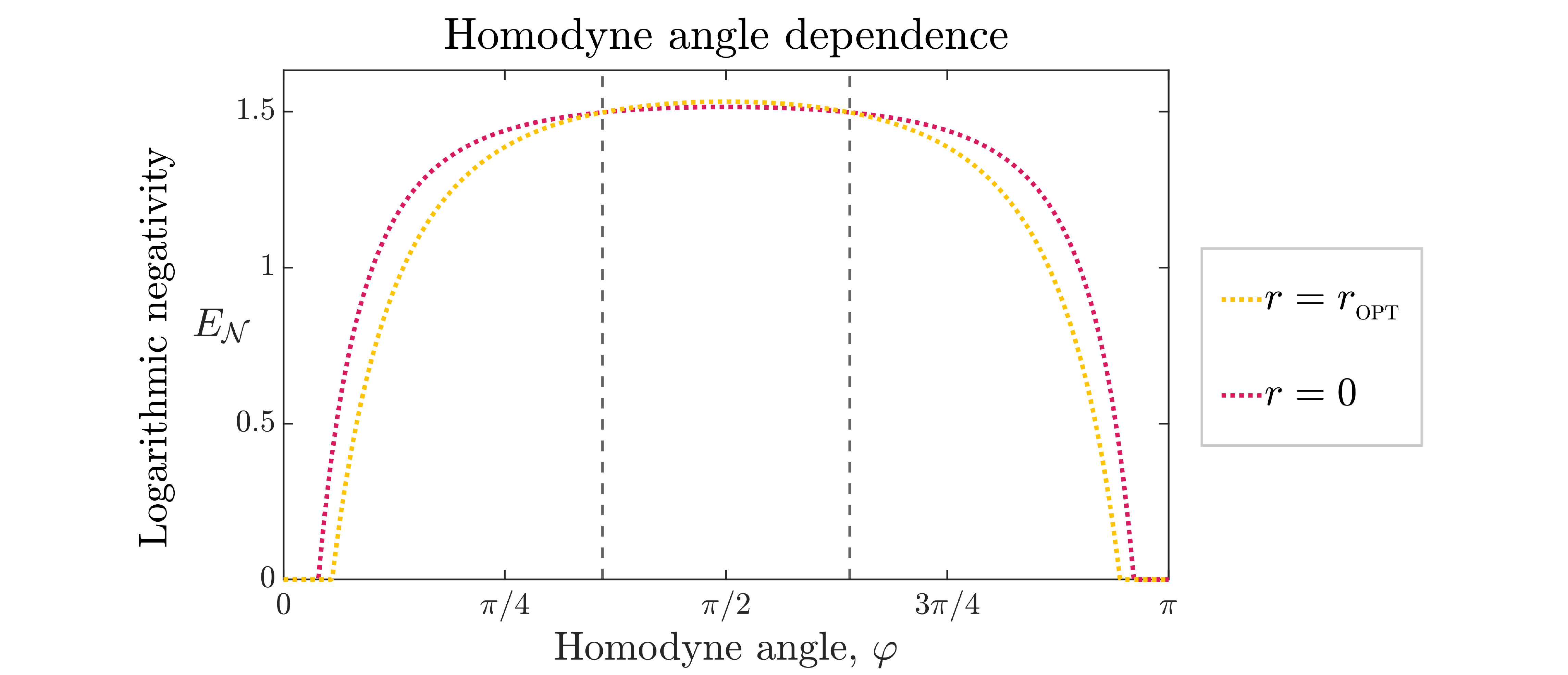}
    \caption{\textbf{Dependence of the logarithmic negativity, with and without optical squeezing, as a function of the optical homodyne angle $\varphi$ in the non-interferometric scheme.} Here, $E_{\mathcal{N}}$ is plotted after the mechanical modes have rotated and decohered through a quarter of a period since entanglement generation. When $r=0$ we set $\chi=3.00$, and when the squeezing parameter is set to be optimal, $r_{\textsc{opt}}=0.15$, we set $\chi=2.97$. These correspond to the values that maximize logarithmic negativity, found in Fig.~\ref{nonintlogneg}. The two curves intersect at $\varphi=1.13$ and $\varphi=2.01$, showing that there is a region around $\varphi=\pi/2$ where squeezing is beneficial to entanglement generation.}
    \label{nonintangles}
\end{figure}
The operator $\Upsilon(\XMo,\XMt,\varphi)$ for the non-interferometric scheme admits a similar polar decomposition as the operator for the interferometric scheme. However, from Fig.~\ref{nonintangles} we see that regardless of whether or not optical squeezers are used, there only exists one angle $\varphi$ that corresponds to maximum entanglement. This occurs at $\varphi=\pi/2$ and corresponds to Bayesian inference of the EPR quadrature. The region between $\varphi=1.13$ and $\varphi=2.01$ is the region where the use of optical squeezers leads to an increase in logarithmic negativity.

\clearpage
%\bibliographystyle{bib1style}
%\bibliography{bib1.bib}

\end{document}